\documentclass[12pt,letterpaper]{article}
\pdfoutput=1

\usepackage{amsmath,amssymb,calc, amsthm,bbm, epsfig,psfrag, mathtools,comment}
\usepackage{dsfont}
\usepackage{mathrsfs}
\usepackage{graphicx, enumerate, enumitem}
\usepackage{float}
\newtheorem{thm}{Theorem}[section]

\usepackage[numbers,sort&compress]{natbib}

\usepackage[dvipsnames]{xcolor}
\usepackage{tikz}
\usepackage{tikz-cd}
\usepackage{tensor}
\newcommand{\ul}{\underline}

\usepackage[utf8]{inputenc} 
\definecolor{azure}{rgb}{0.0, 0.5, 1.0}
\definecolor{darkblue}{rgb}{0.15,0.35,0.7}
\definecolor{reddish}{rgb}{0.65, 0.2, 0.2}
\definecolor{brandeisblue}{rgb}{0.0, 0.44, 1.0}
\definecolor{ceruleanblue}{rgb}{0.16, 0.32, 0.75}
\definecolor{indigo(dye)}{rgb}{0.0, 0.25, 0.42}
\usepackage[linktocpage=true]{hyperref}
\hypersetup{
colorlinks=true,
citecolor=ceruleanblue,
linkcolor=ceruleanblue,
urlcolor=ceruleanblue,
pdfauthor={},
pdftitle={},
pdfsubject={}
}

\newcommand{\overbar}[1]{\mkern 1.5mu\overline{\mkern-1.5mu#1\mkern-1.5mu}\mkern 1.5mu}

\newcommand{\TT}{T\overbar{T}}

\DeclareMathOperator{\tr}{\text{tr}}

\DeclareFontEncoding{LS1}{}{}
\DeclareFontSubstitution{LS1}{stix}{m}{n}
\DeclareSymbolFont{stixsymbols}{LS1}{stixscr}{m}{n}
\SetSymbolFont{stixsymbols}{bold}{LS1}{stixscr}{b}{n}
\DeclareMathSymbol{\kay}{\mathalpha}{stixsymbols}{"6B}
\DeclareMathSymbol{\hay}{\mathalpha}{stixsymbols}{"68}

\makeatletter
\renewcommand\section{\@startsection {section}{1}{\z@}%
                               {-3.5ex \@plus -1ex \@minus -.2ex}
                               {2.3ex \@plus.2ex}%
                               {\normalfont\large\bfseries}}
\renewcommand\subsection{\@startsection{subsection}{2}{\z@}%
                                 {-3.25ex\@plus -1ex \@minus -.2ex}%
                                 {1.5ex \@plus .2ex}%
                                 {\normalfont\bfseries}}
\makeatother

\makeatletter
\newcommand*\bigcdot{\mathpalette\bigcdot@{.5}}
\newcommand*\bigcdot@[2]{\mathbin{\vcenter{\hbox{\scalebox{#2}{$\m@th#1\bullet$}}}}}
\makeatother
\newcommand{\deq}{\stackrel{\bigcdot}{=}}

\newfont{\goth}{ygoth.tfm scaled 1200}                   

\numberwithin{equation}{section}

\newcommand{\del}{\partial}

\newcommand{\ttr}[1]{{\rm tr}\left ( #1\right) }
\newcommand{\rb}[1]{\left( #1\right )}

\allowdisplaybreaks

\begin{document}
\begin{titlepage}
\begin{flushright}
\today
\end{flushright}
\vspace{5mm}

\begin{center}
{\Large \bf 
Higher-Spin Currents and Flows in \\ Auxiliary Field Sigma Models}
\end{center}

\begin{center}

{\bf
Daniele Bielli${}^{a,b}$,
Christian Ferko${}^{c,d}$,
Michele Galli${}^{e}$,\\
Gabriele Tartaglino-Mazzucchelli${}^{e}$
} \\
\vspace{5mm}

\footnotesize{
${}^{a}$
{\it 
High Energy Physics Research Unit, Faculty of Science \\ 
Chulalongkorn University, Bangkok 10330, Thailand
}
 \\~\\
${}^{b}$
{\it 
National Astronomical Research Institute of Thailand \\ 
Don Kaeo, Mae Rim, Chiang Mai 50180, Thailand
}
 \\~\\
${}^{c}$
{\it 
Department of Physics, Northeastern University, Boston, MA 02115, USA
}
 \\~\\
${}^{d}$
{\it 
The NSF Institute for Artificial Intelligence
and Fundamental Interactions
}
  \\~\\
${}^{e}$
{\it 
School of Mathematics and Physics, University of Queensland,
\\
 St Lucia, Brisbane, Queensland 4072, Australia}
}
\vspace{2mm}
~\\
\texttt{d.bielli4@gmail.com,
c.ferko@northeastern.edu,
m.galli@uq.edu.au,
g.tartaglino-mazzucchelli@uq.edu.au
}\\
\vspace{2mm}

\end{center}

\begin{abstract}
\baselineskip=14pt

\noindent We study local, higher-spin conserved currents in integrable $2d$ sigma models that have been deformed via coupling to auxiliary fields. These currents generate integrability-preserving flows introduced by Smirnov and Zamolodchikov. For auxiliary field (AF) deformations of a free boson, we prove that local spin-$n$ currents exist for all $n$ and give recursion relations that characterize Smirnov-Zamolodchikov (SZ) flows driven by these currents. We then show how to construct spin-$2n$ currents in a unified class of auxiliary field sigma models with common structure -- including AF theories based on the principal chiral model (PCM), its non-Abelian T-dual, (bi-)Yang-Baxter deformations of the PCM, and symmetric space models -- for interaction functions of one variable, and describe SZ flows driven by any function of the stress tensor in these cases. Finally, we give perturbative solutions for spin-$3$ SZ flows in any member of our unified class of AF models with underlying $\mathfrak{su}(3)$ algebra. Part of our analysis shows that the class of AF deformations can be extended by allowing the interaction function to depend on a larger set of variables than has previously been considered.

\end{abstract}
\vspace{5mm}

\vfill
\end{titlepage}


\renewcommand{\thefootnote}{\arabic{footnote}}
\setcounter{footnote}{0}

\tableofcontents{}
\vspace{1cm}
\bigskip\hrule

\newpage

\section{Introduction}\label{sec:intro}

The defining characteristic of integrable quantum field theories (IQFTs) in two spacetime dimensions is an infinite tower of conserved quantities.\footnote{We use the phrase ``conserved quantity'' to refer to either a charge $Q$ whose time derivative vanishes, or to a divergence-free current $J$ carrying some number of Lorentz indices.} The presence of such a rich collection of conservation laws allows one to make exact statements about the dynamics of IQFTs, both classically and quantum mechanically. This has led to sustained interest in integrable  theories, which -- far from being mere toy models -- find applications in many areas of theoretical physics, including condensed matter systems, string theory, and holography.

In many of the known $2d$ IQFTs, there are multiple ways to construct infinite towers of conserved quantities, which are qualitatively different and which are each useful in different contexts. Two of the important properties which distinguish various infinite towers are (i) whether the charges are integrals of \emph{local} functions of the fundamental fields in the theory, or \emph{non-local}, and (ii) whether the charges arise from integrated densities
\begin{align}
    Q^{(n)} = \int \mathrm{d} \sigma \, J_\tau^{(n)} \, , \qquad n \in \mathbb{N} \, , 
\end{align}
involving \emph{spin-1} conserved currents $J_\alpha^{(n)}$ which obey $\partial^\alpha J_\alpha^{(n)} = 0$, or if the charges are integrals of densities which sit in a \emph{higher-spin} representation of the Lorentz group.\footnote{In general, non-local charges can also have indefinite or non-integer spin.}

This classification of different conserved charges can be illustrated using the example of the principal chiral model (PCM). Consider a field $g : \Sigma \to G$ from a two-dimensional spacetime $\Sigma$, which we also refer to as the worldsheet, into a Lie group $G$ with Lie algebra $\mathfrak{g}$. Let $j_\pm = g^{-1} \partial_\pm g$ be the pull-back of the Maurer-Cartan form on $G$ to $\Sigma$, where we use light-cone coordinates $\sigma^\pm = \frac{\tau \pm \sigma}{2}$. The PCM is described by the Lagrangian
\begin{align}\label{pcm_lagrangian}
    \mathcal{L}_{\text{PCM}} = - \frac{1}{2} \tr ( j_+ j_- ) \, ,
\end{align}
and the Euler-Lagrange equation associated with the Lagrangian (\ref{pcm_lagrangian}) is
\begin{align}\label{pcm_eom}
    \partial_+ j_- + \partial_- j_+ = 0 \, .
\end{align}
This equation of motion can be recast in a zero-curvature representation by defining a Lie-algebra valued $1$-form called the Lax connection, which takes the form
\begin{align}\label{pcm_lax}
    \mathfrak{L}_{\pm} = \frac{j_\pm}{1 \mp z} \, .
\end{align}
Here $z \in \mathbb{C}$ is called the spectral parameter. Imposing that the Lax (\ref{pcm_lax}) obey the condition
\begin{align}
    \partial_+ \mathfrak{L}_- - \partial_- \mathfrak{L}_+ + [ \mathfrak{L}_+ , \mathfrak{L}_- ] = 0
\end{align}
for all values of the spectral parameter is equivalent to imposing the equation of motion (\ref{pcm_eom}) for the model. A model whose equations of motion can be rewritten as the flatness of such a connection is said to be weakly classically integrable, or just integrable.

Any $2d$ field theory exhibiting weak integrability admits an infinite collection of conserved charges which arise from expanding the monodromy matrix of the Lax connection, $M ( \tau ; z ) = \mathrm{Pexp} \left( - \int_{-\infty}^{\infty} \mathrm{d} \sigma \, \mathcal{L}_\sigma ( \tau, \sigma; z ) \right)$, about a chosen value of the spectral parameter $z$. The conserved charges that are generated by this procedure generically involve multiple nested integrals and commutators of the Lax connection, and are therefore non-local functions of the fields. For instance, in the case of the principal chiral model, expansion of the monodromy matrix about $z = \infty$ gives rise to an infinite tower of conserved charges that can be identified as integrals of temporal components of spin-$1$ conserved currents.

A second infinite tower of conserved quantities in the PCM is given by
\begin{align}\label{local_higher_spin_PCM}
    J_{ \pm n} = \tr ( j_\pm^n) \, .
\end{align}
Using the equation of motion (\ref{pcm_eom}), along with the Maurer-Cartan identity
\begin{align}
    \partial_+ j_- - \partial_- j_+ + [ j_+ , j_- ] = 0 \, ,
\end{align}
which holds by virtue of the definition of $j_\pm$, one can show that each current satisfies
\begin{align}
    \partial_{\mp} J_{ \pm n} = 0 \, .
\end{align}
In contrast to the conserved quantities arising from the Lax connection, the quantities (\ref{local_higher_spin_PCM}) are manifestly local in the fields $j_\pm$, and the conserved current $J_{\pm n}$ carries total Lorentz spin $n$. We therefore classify these as local, higher-spin currents. A systematic discussion of these higher-spin currents appeared in \cite{Evans:1999mj}, where it was shown that there is one such current for each totally symmetric invariant tensor of $G$, of which (\ref{local_higher_spin_PCM}) is just one example.

The existence of two sets of conserved quantities in the PCM -- the non-local ones arising from the Lax connection, and the local higher-spin currents (\ref{local_higher_spin_PCM}) -- is a typical feature which is common to many $2d$ IQFTs. As a general rule, in models with two such towers, each comes with unique properties along with its own advantages and disadvantages. 

\begin{itemize}
    \item[(NL)] The non-local charges generate a quantum group structure, which leads to deep connections with mathematics and implies constraints \cite{LUSCHER19781,Loebbert:2018lxk} on the $S$-matrix in integrable theories. In the case of the PCM, the relevant quantum group is the Yangian \cite{Bernard:1990jw}, a structure which was introduced by Drinfel'd \cite{Drinfeld:1985rx,Drinfeld:1987sy}; see \cite{Loebbert:2016cdm} for a review.

    One advantage of the non-local charges is that they can be constructed via a systematic procedure from the Lax connection, whereas, to the best of our knowledge, no such prescription can be used to universally obtain local higher-spin charges.

    Another benefit of working with the non-local charges is that, because their structure is quite well-studied, one can often import general results and theorems. For instance, to prove that an infinite set of non-local charges are mutually Poisson-commuting, it suffices to show that the Lax exhibits the $r/s$ structure studied by Maillet \cite{MAILLET198654,MAILLET1986401}.

    \item[(L)]\label{advantages_local} Charges which are integrals of local functions of the fields typically take simpler forms which can be easier to manipulate analytically than their non-local counterparts.

    If even two \cite{PARKE1980166} of the local higher-spin charges are non-anomalous, and thus persist in the quantum theory, this severely constrains scattering:  it is sufficient to conclude that the theory exhibits no particle production or annihilation, and that $n \to n$ scattering processes factorize into $2 \to 2$ processes.  Furthermore, the local charges are additive when acting on asymptotic multi-particle states. These properties have made local charges useful in studying $3$-point couplings in affine Toda theories \cite{BRADEN1990689,Arinshtein:1979pb,DOREY1991654}. 

    As we will discuss below, local higher-spin currents can be used to drive integrability-preserving deformations of a given ``seed'' IQFT, generating an infinite family of integrable field theories, due to results of Smirnov and Zamolodchikov \cite{Smirnov:2016lqw}.
\end{itemize}

The complementary properties of these two sets of conserved quantities illustrates the principle that, whenever one is studying an integrable $2d$ theory or family of such theories, it is advantageous to find multiple presentations of the conserved tower in the model.

\enlargethispage{0.2\baselineskip}

In this article, we will be interested in conserved quantities in a larger class of IQFTs which extend the PCM and many related sigma models. The first instances of this class were introduced as deformations of the principal chiral model \cite{Ferko:2024ali}, drawing inspiration from the Ivanov-Zupnik formulation of $4d$ duality-invariant theories of electrodynamics \cite{Ivanov:2002ab,Ivanov:2003uj} and the analyses of \cite{Borsato:2022tmu} and \cite{Ferko:2023wyi}. Soon after, these auxiliary field (AF) deformations were generalized to the non-Abelian T-dual of the PCM (NATD-PCM) \cite{Bielli:2024khq,Bielli:2024ach}, (bi)-Yang-Baxter deformed sigma models \cite{Bielli:2024fnp}, (semi)-symmetric space sigma models (sSSSM) \cite{Bielli:2024oif}, and $\mathbb{Z}_{N}$-coset theories \cite{Cesaro:2024ipq}. The auxiliary field formalism has also been realized via $4d$ Chern-Simons\footnote{The $\TT$ deformation, which is a special case of our auxiliary field deformations, had previously been realized via $4d$ Chern-Simons using a different approach \cite{Py:2022hoa}. See \cite{Lacroix:2021iit} for a review of $4d$ Chern-Simons.} theory \cite{Fukushima:2024nxm} and applied to $2d$ dimensionally reduced gravity \cite{Cesaro:2025msv}. 

In all of these examples, the general strategy is to couple the original field content of an integrable model to a set of auxiliary fields with algebraic equations of motion, while activating prescribed interactions between the auxiliary fields in a way which preserves integrability. For instance, in the case of the auxiliary field deformation of the PCM, one introduces Lie-algebra valued auxiliary fields $v_\pm \in \mathfrak{g}$ and considers the modified Lagrangian
\begin{align}\label{afsm_lag}
    \mathcal{L}_{\text{AFSM}} = \frac{1}{2} \tr ( j_+ j_- ) + \tr ( j_+ v_- + j_- v_+ ) + \tr ( v_+ v_- ) + E ( \nu_2 , \ldots, \nu_N ) \, ,
\end{align}
where
\begin{align}\label{nuk_defn}
    \nu_k = \tr ( v_+^k ) \tr ( v_-^k ) \, , \qquad k = 1 \, , \ldots \, , N \, , 
\end{align}
are a set of $N$ functionally independent scalars constructed from the auxiliary fields, and where $N$ depends on the choice of Lie group $G$. While the generalizations of the auxiliary field couplings to the various other integrable sigma models mentioned above appear more complicated, most of them share a common structure which will be explained in more detail around equation (\ref{AF-general-EOM-for-identity}) below; morally speaking, the mechanism by which integrability is preserved in the more involved cases is the same as for the simplest example (\ref{afsm_lag}). Remarkably, one of the findings of our paper is that the universal implications of (\ref{AF-general-EOM-for-identity}) for integrability hold even when the interaction function in \eqref{afsm_lag} is extended to be an arbitrary Lorentz scalar function of the chiral combinations of auxiliary fields given by $\nu_{\pm k} = \mathrm{tr}(v_{\pm}^k)$.

For each of the known auxiliary field constructions, classical integrability of the deformed model is established by exhibiting a Lax connection which provides a zero-curvature representation of the equations of motion. Therefore, by the general construction mentioned above, one is automatically guaranteed that any such auxiliary field model enjoys an infinite collection of non-local conserved charges, obtained from expanding the monodromy matrix.

However, as we emphasized above, it is useful to find multiple distinct infinite towers of conserved quantities in any integrable model. An important open challenge is to exhibit the local, higher-spin currents in auxiliary field sigma models, generalizing the currents (\ref{local_higher_spin_PCM}) in the PCM. This had not yet been accomplished in previous works, since there is no systematic procedure for extracting such local higher-spin currents from a Lax connection which is applicable to all models. In special cases, such as for the PCM or symmetric space sigma models, the form of the higher-spin currents can be inferred from an Abelianization procedure (see Section 3.1 of \cite{Driezen:2021cpd} or Section 3.8 of \cite{babelon_bernard_talon_2003} for reviews), or by using arguments involving the Maillet $r/s$ structure of the theory \cite{Lacroix:2017isl}. However, these procedures fail even for the standard AFSM, let alone its generalizations to more complicated scenarios.

Thus, the first goal of this paper is to characterize the local higher-spin conserved currents in auxiliary field sigma models. The second main goal is related to the comment, mentioned in point (L) above, that local higher-spin currents may be used to generate integrable deformations of IQFTs. Consider a generic $2d$ field theory which possesses a conserved, totally symmetric spin-$s$ current $J_{\alpha_1 \ldots \alpha_s}$ whose components can be written in light-cone indices as $J_{s \pm}$ and $J_{(s-2) \pm}$. The conservation equation $\partial^{\alpha_1} J_{\alpha_1 \ldots \alpha_s}$ reads
\begin{align}
    \partial_{\mp} J_{\pm s} + \partial_{\pm} J_{\pm(s-2)} = 0 \,.
\end{align}
It was shown in \cite{Smirnov:2016lqw} that, given the existence of such a current and mild assumptions like translation invariance, in any quantum field theory the coincident point limit
\begin{align}\label{OK_defn_intro}
    \mathcal{O}_s ( \sigma ) = \lim_{\sigma' \to \sigma} \left( J_{+s } ( \sigma ) J_{-s} ( \sigma' ) - J_{+(s - 2)} ( \sigma ) J_{-(s - 2)} ( \sigma' ) \right) 
\end{align}
is regular up to divergences which can all be expressed as total derivatives of local operators. These total derivatives vanish when integrated over a spacetime without boundary, so the integrated operator $\mathcal{O}_s$ of equation (\ref{OK_defn_intro}) gives rise to a well-defined deformation of the quantum theory. Remarkably, deforming the model by adding this integrated local operator to the action preserves integrability, and thus -- assuming that the infinitesimally deformed theory still possesses a spin-$s$ conserved current -- one can iterate this flow to generate a one-parameter family of integrable models. We refer to the process of generating such a continuous family of deformed integrable theories as a \emph{Smirnov-Zamolodchikov (SZ) flow}.

The simplest and most famous example of such a flow is the $\TT$ deformation \cite{Zamolodchikov:2004ce,Cavaglia:2016oda}, which corresponds to $s = 2$. Although the $\TT$ flow is well-defined at the quantum level, it is already interesting to study \emph{classical} deformations by this operator, and we will restrict our analysis to classical flows in the remainder of this work. For the case of $\TT$, many results are known about solutions to the classical flow equation for the Lagrangian,
\begin{align}\label{2d_TT_flow}
    \frac{\partial \mathcal{L}_\lambda}{\partial \lambda} = \frac{1}{4} \left( T^{(\lambda) \alpha \beta} T^{(\lambda)}_{\alpha \beta} - \left( \tensor{T}{^{(\lambda)}^{\alpha}_{\alpha}} \right)^2 \right) \, ,
\end{align}
where $T_{\alpha \beta}^{(\lambda)}$ is the stress tensor associated with the Lagrangian $\mathcal{L}_\lambda$. The solution to (\ref{2d_TT_flow}) with an initial condition given by the Lagrangian of $2d$ free bosons is the gauge-fixed Nambu-Goto Lagrangian \cite{Cavaglia:2016oda,Bonelli:2018kik}. Similar flow equations have been studied in other $2d$ models such as gauge theories \cite{Conti:2018jho,Brennan:2019azg} and chiral boson models \cite{Ouyang:2020rpq,Chakrabarti:2020dhv,Ebert:2024zwv,Babaei-Aghbolagh:2025lko}, in four dimensions \cite{Conti:2018jho,Ferko:2022iru,Ferko:2023ruw,Ferko:2023wyi,Ferko:2024yua,Babaei-Aghbolagh:2024uqp}, three dimensions \cite{Ferko:2023sps}, six dimensions \cite{Ferko:2024zth,Kuzenko:2025jgk}, in quantum mechanics \cite{Gross:2019ach,Gross:2019uxi,Chakraborty:2020xwo,Ferko:2023ozb,Giordano:2023wgx,Ferko:2023iha}, in cases with supersymmetry \cite{Baggio:2018rpv,Chang:2018dge,Chang:2019kiu,Jiang:2019hux,Coleman:2019dvf,Ferko:2019oyv,Ferko:2021loo,Lee:2021iut,Lee:2023uxj}, for sequential $\TT$-like flows \cite{Ferko:2022dpg}, and for related deformations like root-$\TT$ \cite{Ferko:2022cix,Conti:2022egv,LukeMartin:2023hdw,Babaei-Aghbolagh:2022uij,Babaei-Aghbolagh:2022leo,Babaei-Aghbolagh:2025cni}. The solution of these flow equations often involves mathematical techniques like the method of characteristics \cite{Hou:2022csf} or geometrical approaches \cite{Conti:2018tca,Conti:2022egv,Morone:2024ffm,Tsolakidis:2024wut,Babaei-Aghbolagh:2024hti,Brizio:2024arr,Ferko:2024yhc,Morone:2024sdg,Hao:2024stt}, and has led to new insights in several areas, such as connections to brane physics \cite{Blair:2024aqz} and ModMax-type theories \cite{Bandos:2020jsw,Bandos:2020hgy}. See the reviews \cite{Jiang:2019epa,He:2025ppz} for more details about $\TT$ flows.

Despite the wealth of results about classical stress tensor flows, 
and the study of deformations of the free boson in \cite{Rosenhaus:2019utc}, 
almost nothing is known about the solutions to  Smirnov-Zamolodchikov Lagrangian flows driven by operators $\mathcal{O}_s$ with $s \geq 3$.\footnote{Somewhat different, and Lorentz-breaking, flows driven by operators of the schematic form $T J_s$, involving products of the stress tensor with another spin-$s$ current for $s \geq 3$, have been studied in \cite{Conti:2019dxg}.}
This brings us to the second main goal of this article. One might hope that the auxiliary field formalism could help in studying classical SZ flows, since it was pointed out in \cite{Bielli:2024ach} that the leading-order effect of an auxiliary field deformation by an interaction function $E  = \lambda \nu_s$ is precisely to implement a spin-$s$ Smirnov-Zamolodchikov deformation to first order in $\lambda$. In this work, we extend this observation beyond leading order. We will see that, in many cases, it is possible to prove the existence of all-orders solutions to higher-spin Smirnov-Zamolodchikov flows using the auxiliary field formalism, and to characterize these solutions perturbatively. We hope that these results represent a first step towards a deeper understanding of these higher-spin deformations, which -- if the considerable insight that has been gleaned from the solution to $\TT$-like spin-$2$ flows is any indication -- could open the door to an entirely new set of results that may teach us a great deal about deformations of quantum field theories.

The layout of this paper is as follows. In Section \ref{section:review}, we review aspects of auxiliary field sigma models and how the higher-spin deformations introduced in \cite{Bielli:2024ach} can be further extended while still preserving the integrability of the deformed theory. As an illustrative example, Section \ref{sec:free_boson} pursues the two primary goals of this paper in the simplified setting of a single free boson, both characterizing the higher-spin conserved currents and studying the classical Smirnov-Zamolodchikov flows driven by these currents. In Section \ref{sec:spin-2k-of-AFSM}, we show that spin-$2n$ conserved currents can be constructed in a large class of auxiliary field models satisfying certain assumptions (\ref{AF-general-EOM-for-identity}) by reducing the problem to one which resembles that of the free boson. The results of this section allow even-spin currents, and the associated SZ flows, to be studied in many cases, including the standard AFSM and its non-Abelian T-dual, (bi-)Yang-Baxter AF models, and symmetric space auxiliary field models.
Section \ref{sec:su3} studies the case of auxiliary field models with $\mathfrak{g} = \mathfrak{su} ( 3 )$ and identifies the spin-$3$ currents for models obeying a Smirnov-Zamolodchikov flow equation which is itself driven by these spin-$3$ currents. Section \ref{sec:conclusion} summarizes our results and presents directions for future research. We have collected several ancillary calculations and identities in Appendices \ref{appendix:ODEs}, \ref{appendix:identity}, and \ref{app:su3der}.

\section{Review of auxiliary field (AF) deformations}\label{section:review}

In this section we briefly recall basic facts about auxiliary field deformations of $2d$ integrable sigma models.\footnote{For more general reviews of integrable sigma models in $2d$, as well as some of the integrable deformations of these models which we will mention in this work, see \cite{Zarembo:2017muf,Orlando:2019his,Seibold:2020ouf,Klimcik:2021bjy,Hoare:2021dix,Borsato:2023dis}.} As we mentioned in the introduction, while the original AF formalism \cite{Ferko:2024ali} was shown to include deformations by arbitrary functions of the stress tensor, successive works \cite{Bielli:2024ach,Bielli:2024fnp,Bielli:2024oif} highlighted the possibility of extending the construction to deformations by higher-spin conserved currents, pointing out a connection with the deformations of Smirnov and Zamolodchikov \cite{Smirnov:2016lqw}, whose analysis is the second main goal of this work. Although we have already introduced the auxiliary field deformation of the PCM in equation (\ref{afsm_lag}), we will now describe how such AF deformations are implemented more generally.

The AF deformations of a $2d$ sigma model are constructed by introducing a set of auxiliary fields $v$, which couple to the fundamental fields of the theory, and to themselves, in such a way that the undeformed model is recovered in the limit of trivial $v$-interactions. All such sigma models exhibit the following structure:
\begin{equation}
S^{\text{E}}[\phi,v] := \int_{\Sigma}\mathrm{d}\sigma^{+}\mathrm{d}\sigma^{-} \, \mathcal{L}^{\text{E}}(\phi,v) 
\qquad \text{with} \qquad
\mathcal{L}^{\text{E}}(\phi,v) := \mathcal{L}(\phi,v) + E(v) \, ,
\end{equation}
where $\Sigma$ denotes a flat 2d Lorentzian worldsheet with lightcone coordinates $\sigma^{\pm}$ and
\begin{itemize}
\item $\phi$ schematically denotes the fundamental fields of the theory -- typically interpreted as coordinates on some background $\mathcal{M}$ and regarded as maps $\phi: \Sigma \rightarrow \mathcal{M}$.
\item $v$ denotes the auxiliary fields -- Lie algebra valued as in \cite{Ferko:2024ali,Bielli:2024khq,Bielli:2024ach,Bielli:2024fnp,Bielli:2024oif}, naturally defined on $\Sigma$, and transforming with the fundamental fields under the isometries of $\mathcal{M}$ \cite{Bielli:2024khq}.
\item $E(v)$ is an unspecified function encoding the interaction of auxiliary fields. In \cite{Bielli:2024ach,Bielli:2024fnp,Bielli:2024oif} the $v$-dependence was restricted to the following set of Lorentz invariant combinations
\begin{equation}\label{nu-k-variables-definition}
E(v):=E(\nu_{2},...,\nu_{N}) 
\qquad
\text{with} 
\qquad \nu_{n}:=\mathrm{tr}(v_{+}^n)\mathrm{tr}(v_{-}^n) \qquad \forall \,  n=2,...,N \, ,
\end{equation}
where $N$ is a large enough integer to guarantee having a complete set of algebraic structures describing completely symmetric invariant tensors -- for instance, $N$ could be the rank of the Lie algebra.
However, as it will become clear in later sections, it is crucial that $E(v)$ can be extended to the following larger set of interaction functions
\begin{equation}\label{newE}
E(v):=E(\nu_{+2},...,\nu_{+N},\nu_{-2},...,\nu_{-N}) 
\quad \text{with} \quad
\nu_{\pm n}:=\mathrm{tr}(v_{\pm}^n) 
\quad \forall \, n\in \{2,...N\} \, .
\end{equation}
It is indeed straightforward to check that a key sufficient condition used in our previous works 
\cite{Ferko:2024ali,Bielli:2024khq,Bielli:2024ach,Bielli:2024fnp,Bielli:2024oif} to show that classical integrability is preserved by all these deformations, namely the equation
\begin{equation}\label{deltapm_in_review_sec}
[\Delta_{\pm}, v_{\mp}] \deq 0
\qquad \text{with} \qquad 
\Delta_{\pm}:=\delta_{v_{\mp}}E(\nu_{2},...,\nu_{N}) \, ,
\end{equation}
remains unaffected by the choice \eqref{newE}. More explicitly, one now finds
\begin{equation}\label{Delta-enlarged-ansatz}
\Delta_{\pm}:=\delta_{v_{\mp}}E(\nu_{+2},...,\nu_{+N},\nu_{-2},...,\nu_{-N})=\sum_{n=2}^{N} n \frac{\partial E}{\partial \nu_{\mp n}} v_{\mp}^{A_{1}}...v_{\mp}^{A_{n-1}}d_{A_{1}...A_{n-1}}{}^{B}T_{B} \, .
\end{equation}
Each term in the sum only differs by a scalar factor $\mathrm{tr}(v_{\pm}^k)$, from the ones obtained by varying \eqref{nu-k-variables-definition}, and this cannot affect the commmutator \eqref{deltapm_in_review_sec}. In \eqref{Delta-enlarged-ansatz} we used
\begin{equation}\label{d-symmetric-tensor}
d_{A_{1}...A_{n-1}}{}^{B}:=\mathrm{tr}(T_{(A_{1}}...T_{A_{n-1}}T_{C)})\gamma^{CB}
\qquad \text{and} \qquad
\gamma_{AB}:=\mathrm{tr}(T_{A}T_{B}) \, ,
\end{equation}
$\gamma_{AB}$ being the Cartan-Killing form and $\gamma^{AB}$ its inverse. As in previous works, the $\deq$ symbol denotes equality on-shell for the auxiliary fields. Within the extension \eqref{newE}, the requirement of Lorentz invariance translates into the homogeneity condition
\begin{equation}\label{homogeneity-condition}
\sum_{n=2}^{N} \! n \, \nu_{+n} \, \frac{\partial E}{\partial\nu_{+n}} \!=\! \sum_{n=2}^{N} \! n \, \nu_{-n} \, \frac{\partial E}{\partial\nu_{-n}} 
\qquad \text{or equivalently} \qquad
\mathrm{tr}(\Delta_{-}v_{+})\!=\!\mathrm{tr}(\Delta_{+}v_{-}) \, .
\end{equation}
\end{itemize}

The Lagrangian $\mathcal{L}^{\text{E}}(\phi,v)$ is constructed in such a way that for trivial interaction functions $E=0$, it is very simple to integrate out the auxiliary fields and recover the Lagrangian $\mathcal{L}(\phi)$ of the original undeformed theory, also known as \textit{seed theory}. On the other hand, integrating out the auxiliary fields in the presence of non-trivial interaction functions (whenever this is possible) leads to generically complicated deformations $\mathcal{L}_{\mathcal{O}}(\phi)$ of the seed theory. For specific choices of $E$, these have been shown to take particularly nice forms, including $T\bar{T}$, root-$T\bar{T}$ and deformations by operators which are constructed out of higher-spin conserved currents. The latter case is particularly relevant here: choosing 
\begin{equation}\label{E-leading-order}
E(\nu_{2},...,\nu_{N}) := \lambda_{n} \nu_{n} \qquad \text{for any} \qquad n \in\{2,...,N\} 
\,,
\end{equation}
leads, at least to leading order in the spin-$n$ deformation parameter $\lambda_{n}$, to deformations 
\begin{equation}\label{L-deformation-leading-order}
\mathcal{L}_{\mathcal{O}_{n}}(\phi) \simeq \mathcal{L}(\phi)+\lambda_{n}\mathcal{O}_{n} \qquad \text{with} \qquad \mathcal{O}_{n}:=\mathcal{J}_{+n}\mathcal{J}_{-n} \, ,
\end{equation}
where $\mathcal{J}_{\pm n}$ are spin-$n$ currents of the seed theory which are conserved on-shell: $\partial_{\pm}\mathcal{J}_{\mp n}=0$. In turn, these deformations drive a flow of the seed Lagrangian which takes the form
\begin{equation}
\frac{\partial S_{\mathcal{O}_{n}}[\phi]}{\partial \lambda_{n}} \simeq \int_{\Sigma}\mathrm{d}\sigma^{+}\mathrm{d}\sigma^{-} \, \mathcal{O}_{n} \, ,
\end{equation}
and can in principle become arbitrarily complicated beyond the leading order in $\lambda_{k}$.

In this work, we will focus our attention on the following two tasks:
\begin{itemize}
\item Constructing higher-spin conserved currents $\tau_{\pm n}$ and $\theta_{\pm (n-2)}$ satisfying
\begin{equation}\label{conservation}
\partial_{\pm}\tau_{\mp n} + \partial_{\mp}\theta_{\mp(n-2)} = 0 \, ,
\end{equation}
for auxiliary field models with generic interaction functions.
\item Identifying interaction functions which, beyond leading order in the deformation parameter $\lambda_{n}$, define models that obey a Smirnov-Zamolodchikov-type flow 
\begin{equation}\label{flow}
\frac{\partial S_{\mathcal{O}_{n}}[\phi]}{\partial \lambda_{n}} \simeq \int_{\Sigma}\mathrm{d}\sigma^{+}\mathrm{d}\sigma^{-} \, \mathcal{O}_{n}
\qquad
\text{with} 
\qquad 
\mathcal{O}_{n}:= \tau_{+n}\tau_{-n}-\theta_{+(n-2)}\theta_{-(n-2)} \, .
\end{equation}
\end{itemize}
It is convenient to make here a couple of quick dimensional analysis remarks, which will hold true for any AF sigma model. We will work in units such that any action $S^{\text{E}}[\phi,v]$ is dimensionless and consequently any Lagrangian must have units of inverse-length squared:
\begin{equation}\label{dim-analysis1}
[S^{\text{E}}]=0
\qquad , \qquad
[\sigma^{\pm}] = -1
\qquad , \qquad
[\mathcal{L}^{\text{E}}]=2 
\qquad , \qquad
[E]=2 \, .
\end{equation}
Every model exhibits a kinetic-like term for the auxiliary fields which fixes their dimension
\begin{equation}\label{dim-analysis2}
\mathrm{tr} ( v_{+}v_{-} )  \subset \mathcal{L}(\phi,v) 
\qquad \Rightarrow \qquad
[v_{\pm}] = 1 \, .
\end{equation}
In turn, this fixes the dimension of $\nu_{n}$ \eqref{nu-k-variables-definition}, $\lambda_{n}$ \eqref{E-leading-order}, $\mathcal{O}_{n}$ \eqref{L-deformation-leading-order} and $\tau_{\pm n},\theta_{\pm(n-2)}$ \eqref{flow} as
\begin{equation}\label{dim-analysis3}
[\nu_{n}] = 2n
\,\,\,\,\, , \,\,\,\,\,
[\lambda_{n}] = 2-2n 
\,\,\,\,\, , \,\,\,\,\,
[\mathcal{O}_{n}] = 2n
\,\,\,\,\, , \,\,\,\,\,
[\tau_{\pm n}] = n
\,\,\,\,\, , \,\,\,\,\,
[\theta_{\pm(n-2)}] = n \, .
\end{equation}

\section{An instructive toy model -- the AF free boson}\label{sec:free_boson}

We begin our discussion with the simplest $2d$ sigma model: the single free boson. Despite its elementary structure this model exhibits interesting features, which will appear again at later stages, and for this reason represents a useful playground that can be used as a guideline for more complicated settings. Higher-spin currents for this theory were also constructed in \cite{Rosenhaus:2019utc} and contact with such results will be made at the end of this section.\footnote{See also \cite{Lindwasser:2024qyh,Lindwasser:2025slu} for recent discussions of another large class of integrable deformations of the theory of a single free boson, which does not include the auxiliary field deformations considered here.}

The Lagrangian for a free boson deformed by auxiliary field couplings takes the form
\begin{equation}\label{Lagrangian-free-boson}
\mathcal{L}_{\varphi}^{\text{E}} = \frac{1}{2}\partial_{+}\varphi \partial_{-} \varphi+v_{+}v_{-}+v_{+}\partial_{-}\varphi + \partial_{+}\varphi v_{-} + E(\nu) \qquad \text{with} \qquad \nu:=v_{+}v_{-} \, .
\end{equation}
Notice that while for more complicated models one has several independent variables $\nu_{n}$, defined in \eqref{nu-k-variables-definition}, which allow to deform the theory, for the free boson there is only one independent variable $\nu$, since any higher $\nu_{n}$ would simply be rewritten as $\nu_{n}=\nu^{n}$. For this reason, a generic function $E(\nu)$ is in this case sufficient to take into account all possible types of deformations. The EOMs for the model read
\begin{equation}\label{EOM}
\begin{aligned}
\delta_{\varphi}\mathcal{L}_{\varphi}^{\text{E}}&=-\partial_{+}\partial_{-}\varphi-\partial_{-}v_{+}-\partial_{+}v_{-} \equiv 0 \, ,
\\
\delta_{v_{+}}\mathcal{L}_{\varphi}^{\text{E}}&=v_{-}+\partial_{-}\varphi+E'v_{-} \deq 0 \, ,
\\
\delta_{v_{-}}\mathcal{L}_{\varphi}^{\text{E}}&=v_{+}+\partial_{+}\varphi+E'v_{+} \deq 0 \, ,
\end{aligned}
\end{equation}
and for $E=0$ the last two give $v_{\pm}\deq-\partial_{\pm}\varphi$, allowing to recover the undeformed Lagrangian
\begin{equation}
\mathcal{L}_{\varphi}^{\text{E}=0}\deq-\frac{1}{2}\partial_{+}\varphi \partial_{-} \varphi =:\mathcal{L}_{\varphi} \, .
\end{equation}

Section \ref{sec:intro} described two main goals of this work. The first is to characterize the local higher-spin conserved currents in auxiliary field sigma models. This question is completely agnostic as to the choice of interaction function; given \emph{any} such function $E$, there should exist a tower of local higher-spin conserved currents. For the case of a single free boson which is our focus in this section, this goal will be achieved in section \ref{subsec:free_boson_currents}, since (as we will see) the system of ordinary differential equations (\ref{system-original-form}) always admits solutions for $f$ and $g$ which provide us with a local spin-$n$ current whose components are $\tau_{\pm n}$ and $\theta_{\pm ( n - 2 )}$.

The second objective is to describe Smirnov-Zamolodchikov flows driven by combinations of higher-spin currents. Unlike the first goal, this question is \emph{not} agnostic to the choice of interaction function: indeed, we are explicitly interested in finding the function $E$ which obeys a flow equation \eqref{flow}. This additional assumption about the interaction function imposes further constraints. What we mean by a solution to this second problem is a characterization of both the interaction function $E$, and the conserved higher-spin currents $\tau_{\pm n}$ and $\theta_{\pm ( n - 2 )}$, as a function of $\lambda_n$, such that this collection of data obeys \eqref{flow}. This second goal will be accomplished in Section \ref{subsec:perturbative-analysis-free-boson} for deformations of a free boson.

\subsection{Higher-spin currents}\label{subsec:free_boson_currents}
\noindent
We begin with the first goal, which is to characterize the local higher-spin currents in an auxiliary field model for a single free boson with generic interaction function. Given the simplicity of the model, one can make a general ansatz for the higher-spin currents
\begin{equation}\label{free-bosono-higher-spin-currents-guess}
\tau_{\pm n} := f(\nu) v_{\pm}^{n}
\qquad ,\qquad 
\theta_{\pm(n-2)} := g(\nu) v_{\pm}^{n-2} \, .
\end{equation}
The next step is trying to determine the most general form of $f(\nu),g(\nu)$ satisfying the conservation equation \eqref{conservation}, which takes the following explicit form:
\begin{equation}\label{conservation-intermediate-step1}
\begin{aligned}
\partial_-{\tau_{+n}}+\partial_{+}\theta_{+(n-2)}&=[\nu f'+nf]v_{+}^{n-1}\partial_{-}v_{+}+g'v_{+}^{n-1}\partial_{+}v_{-}+
\\
&  + [\nu g'+(n-2)g]\frac{v_{+}^{n-1}}{\nu^2} v_{-}^2\partial_{+}v_{+}+f'v_{+}^{n-1}v_{+}^2\partial_{-}v_{-} \, .
\end{aligned}
\end{equation}
One can immediately notice the existence of four main contributions, respectively proportional to $\partial_{\pm}v_{+}$ and $\partial_{\pm}v_{-}$, whose coefficients should vanish for the conservation equation to be satisfied. While these structures are a priori independent, the EOM \eqref{EOM} clearly establish relations among them and since any conserved current is meant to be so when going on-shell, one can use \eqref{EOM} to connect the various terms in \eqref{conservation-intermediate-step1}. Differentiating the second and third equation in \eqref{EOM} by, respectively, $\partial_{+}$ and $\partial_{-}$ one finds
\begin{equation}
\begin{aligned}
(1+E'+\nu E'')\partial_{+}v_{-}+E''v_{-}^2\partial_{+}v_{+}+\partial_{+}\partial_{-}\varphi&=0 \, , 
\\
(1+E'+\nu E'')\partial_{-}v_{+}+E''v_{+}^2\partial_{-}v_{-}+\partial_{+}\partial_{-}\varphi&=0 \, , 
\end{aligned}
\end{equation}
and substituting $\partial_{+}\partial_{-}\varphi$ from the first equation in \eqref{EOM} the following relations are obtained
\begin{equation}\label{identities}
\begin{aligned}
E'' v_{-}^2 \partial_{+}v_{+} &= (\partial_{-}v_{+}-(\nu E')'\partial_{+}v_{-}) \, , 
\\
E'' v_{+}^2 \partial_{-}v_{-} &= (\partial_{+}v_{-}-(\nu E')'\partial_{-}v_{+}) \, .
\end{aligned}
\end{equation}
The latter can now be exploited to express two of the four structures in \eqref{conservation-intermediate-step1} in terms of the remaining ones. Notice that for choices of interaction functions satisfying $E''=0$, there would be no relation between the four structures and consistency of the on-shell relations \eqref{identities} would in fact enforce $E'=1$.
For this reason we will from now on assume that $E''\neq 0$: under this assumption one can immediately exploit \eqref{identities} to rewrite \eqref{conservation-intermediate-step1} as 
\begin{equation}\label{conservation-intermediate-step}
\begin{aligned}
\partial_-{\tau_{+n}}+\partial_{+}\theta_{+(n-2)}& = v_{+}^{n-1}\partial_{-}v_{+}\Bigl[ \nu f' + nf + \frac{\nu g' + (n-2)g}{\nu^2 E''} - \frac{f'(\nu E')'}{E''}\Bigr]+
\\
& +v_{+}^{n-1}\partial_{+}v_{-}\Bigl[ g'- (\nu E')' \frac{\nu g' + (n-2)g}{\nu^2 E''} + \frac{f'}{E''}   \Bigr] \, ,
\end{aligned}
\end{equation}
such that requiring the whole expression to vanish leads to the conditions\footnote{Notice that considering the equation $\partial_{+}\tau_{-n}+\partial_{-}\theta_{-(n-2)} = 0$, instead of $\partial_{-}\tau_{+n}+\partial_{+}\theta_{+(n-2)} = 0$, one would end up with the same set of conditions on $f,g$. Similarly, solving the relations \eqref{identities} for $\partial_{-}v_{+}$ and $\partial_{+}v_{-}$ and substituting the result in \eqref{conservation-intermediate-step1}, the equations remain unchanged.}
\begin{equation}\label{system-original-form}
\begin{aligned}
0&=f' + nf + \frac{\nu g' + (n-2)g}{\nu^2 E''} - \frac{f'(\nu E')'}{E''} 
\, ,
\\
0&=g'- (\nu E')' \frac{\nu g' + (n-2)g}{\nu^2 E''} + \frac{f'}{E''} \, .
\end{aligned}
\end{equation}
The above set of conditions can be regarded as a system of coupled 1st order ODEs for the two functions $f$ and $g$, such that a solution can be found given any choice of interaction function $E$ satisfying suitable conditions for the existence and uniqueness theorem to hold. Formally, this represents a solution to the first problem we set out to study, since we have proven that local higher-spin currents always exist and can be obtained by solving (\ref{system-original-form}).

Given the simplicity of the model and the structure of the equations, one may in principle hope to be able to find a closed-form expression for $f$ and $g$ in terms of $E$ and its derivatives; however, as we will discuss in the next subsection, it turns out that no simple solution of this form can be found, and one is forced to resort to other techniques. Despite these difficulties, the system \eqref{system-original-form} exhibits a very nice structure and can be rewritten in various intriguing forms, which we briefly discuss in appendix \ref{appendix:ODEs}.

\subsubsection*{\ul{\it No-go theorem for closed-form solution}}

In the case of the spin-$2$ conserved current, which is just the stress tensor, one finds a simple closed-form expression in terms of the interaction function $E$ and its derivatives:
\begin{align}\label{stress_tensor_boson}
    T_{\pm \pm} = \left( - 1 + E^{\prime 2} \right) v_{\pm} v_{\pm} \, , \qquad T_{+-} = 2 \left( E' \nu - E  \right) \, .
\end{align}
We might na\"ively expect that a similar closed-form solution exists for all of the higher-spin currents, i.e. that the system of equations (\ref{system-original-form}) admits a solution for $f$ and $g$ which can be written in terms of $E$ and finitely many of its derivatives. Unfortunately, although a solution to these equations does exist generically, it does not admit such a simple form. This result is encoded in the following theorem.

\begin{thm}\label{nogo_thm}
Consider an auxiliary field sigma model which describes a deformation of a single free boson and is characterized by an interaction function $E = E ( \nu )$. Suppose that we make an ansatz for higher-spin conserved currents in the model of the form
\begin{align}\label{theorem_ansatz}
    \tau_{\pm n} = F ( E, E', E'' , \ldots , E^{(N)} ) v_{\pm}^n \, , \qquad \theta_{\pm ( n - 2 ) } = G ( E, E', E'', \ldots, E^{(N)} ) v_{\pm}^{n-1} v_{\mp} \, ,
\end{align}
where the functions $F$ and $G$ depend only on dimensionless combinations involving the interaction function and finitely many of its derivatives. Here $N$ is a positive integer and $E^{(j)}$ denotes the $j$-th derivative $\frac{d^j E}{d \nu^j}$.Then the system of equations arising from the conservation condition $\partial_{\mp} \tau_{\pm n} + \partial_{\pm} \theta_{\pm ( n - 2 ) } = 0$ admits no non-trivial solution for generic $E$ and $n \geq 3$.
\end{thm}

Let us first remark on the form of the ansatz (\ref{theorem_ansatz}), which differs from equation (\ref{system-original-form}) only by the introduction of a factor of $v_\mp$ in $\theta_{\pm ( n - 2 )}$. Using the equations of motion, any expression involving the fields $v_\pm$ and $j_\pm = \partial_\pm \phi$ in the model can be expressed entirely in terms of $v_\pm$. Therefore, the most general expression for a spin-$k$ quantity in the theory is $J_{\pm k} = f ( \nu ) v_\pm^k$. Since $v_\pm$ has mass dimension $1$, if the quantity $J_{\pm k}$ is to have mass dimension $k$ (as is appropriate for a spin-$k$ conserved current generalizing the undeformed expressions $\tr ( j_\pm^k )$), then the quantity $f ( \nu )$ must be dimensionless. Note that, by absorbing factors of $\nu$ into the function $f$, any such quantity can also be written as $J_{ \pm k} = \tilde{f} ( \nu ) v_{\pm}^{m + k} v_{\mp}^m$ for any integer $m$; we have chosen to write $\theta_{\pm ( n - 2) }$ in this form with $m = 1$ for convenience. Therefore, the ansatz (\ref{theorem_ansatz}) represents the most general expressions for two quantities of spins $n$ and $(n-2)$, with appropriate mass dimensions, in the theory.

\begin{proof}

We first enumerate the independent dimensionless quantities that can be constructed from $\nu$, $E$, and the first $N$ derivatives of $E$. Since $E$ must have mass dimension $2$ and $v_{\pm}$ each have mass dimension $1$, the following quantities are dimensionless Lorentz scalars:
\begin{align}
    X_0 = \frac{1}{\nu} E ( \nu ) E' ( \nu ) \quad , \; X_1 = E' ( \nu ) \quad , \; X_2 = \nu E'' ( \nu ) \quad , \; \ldots \quad , \; X_N = \nu^{N-1} E^{(N)} ( \nu ) \, .
\end{align}
Any other dimensionless scalar which depends only on $E$ and finitely many of its derivatives can be expressed in terms of the variables $X_i$. For instance, $Y = \frac{E ( \nu )}{\nu}$ is dimensionless and can be written as $Y = \frac{X_0}{X_1}$. We therefore refine our ansatz (\ref{theorem_ansatz}) for the currents to
\begin{align}\label{theorem_ansatz_refined}
    \tau_{\pm n} = F ( X_0, X_1, \ldots, X_N ) v_{\pm}^n \, , \qquad \theta_{\pm ( n - 2 ) } = G ( X_0, X_1 , \ldots, X_N ) v_{\pm}^{n-1} v_{\mp} \, .
\end{align}
We now consider the terms that appear in the conservation equation
\begin{align}\label{conservation_thm}
    \partial_{\mp} \tau_{\pm n} + \partial_{\pm} \theta_{\pm ( n - 2 ) } = 0 \, .
\end{align}
Upon taking these derivatives, we will generate terms of the form $\partial_{\pm} v_{\mp}$ and $\partial_{\pm} v_{\pm}$. In principle, this gives four separate derivative quantities that can appear. However, we can eliminate two of these terms using the equations of motion of the model and the Maurer-Cartan identity, much like the procedure carried out around equation (\ref{identities}) above. We differentiate the equation of motion
\begin{align}
    v_{\pm} + j_{\pm} + E' v_{\pm} = 0 \, ,
\end{align}
where we write $j_\pm = \partial_\pm \phi$, and solve for two of the derivatives of $v_{\pm}$, eliminating derivatives of $j_\pm$ by using the conservation equation
\begin{align}
    \partial_+ \left( j_- + 2 v_- \right) + \partial_- \left( j_+ + 2 v_+ \right) = 0 \, ,
\end{align}
as well as the Maurer-Cartan identity (or symmetry of mixed second derivatives)
\begin{align}
    \partial_+ j_- = \partial_- j_+ \, .
\end{align}
The result of this procedure is the pair of equations
\begin{equation}\label{elim_derivs_thm}
\begin{aligned}
    \partial_- v_- &= \frac{\partial_+ v_- - \left( E' + \nu E'' \right) \partial_- v_+}{v_+^2 E''} \, ,  \\
    \partial_+ v_+ &= \frac{\partial_- v_+ - \left( E' + \nu E'' \right) \partial_+ v_-}{v_-^2 E''} \, .
\end{aligned}
\end{equation}
We then substitute the relations (\ref{elim_derivs_thm}) into the conservation equation (\ref{conservation_thm}), which eliminates two of the four derivatives of $v_\pm$ that can appear. For the moment, we assume $N \geq 2$ and we focus on the term in the resulting equation which is proportional to the highest derivative of the interaction function, which takes the form
\begin{equation}\label{highest_term_thm}
\begin{aligned}
    \left( \partial_{\mp} \tau_{\pm n} + \partial_{\pm} \theta_{\pm ( n - 2 ) } \right) \Big\vert_{E^{(N+1)}} &\sim E^{(N+1)} \left( \frac{\partial G}{\partial X_N} - X_1 \frac{\partial F}{\partial X_N} \right) \partial_- v_+ \\
    &\qquad + E^{(N+1)} \left( \frac{\partial F}{\partial X_N} - X_1 \frac{\partial G}{\partial X_N} \right) \partial_+ v_- \, ,
\end{aligned}
\end{equation}
where the notation $\Big\vert_{E^{(N+1)}}$ means that we have extracted the term proportional to $E^{(N+1)}$. There is no other term in the conservation equation which involves $E^{(N+1)}$, and the functions $F$ and $G$ do not depend on this quantity by assumption, so in order to satisfy the conservation equation, both of the terms in (\ref{highest_term_thm}) must vanish identically:
\begin{align}
    \frac{\partial G}{\partial X_N} - X_1 \frac{\partial F}{\partial X_N} = 0 = \frac{\partial F}{\partial X_N} - X_1 \frac{\partial G}{\partial X_N} \, .
\end{align}
These two equations imply that
\begin{align}
    \frac{\partial F}{\partial X_N} = 0 = \frac{\partial G}{\partial X_N} \, ,
\end{align}
so consistency requires that the functions $F$ and $G$ are actually independent of the combination $X_N$. But then one can repeat the argument above, collecting terms that are proportional to $E^{(N)}$ in the conservation equation and concluding that the functions $F$ and $G$ are independent of $X_{N-1}$. We continue in this way, eliminating dependence of $F$ and $G$ on each of the variables $X_i$, until we arrive at $i = 1$ and we are forced to assume
\begin{align}
    F = F ( X_0, X_1 ) \, , \qquad G = G ( X_0, X_1 ) \, .
\end{align}
In this case, the conservation equation (\ref{conservation_thm}) takes the form
\begin{equation}\label{four_pde_eqn_thm}
\begin{aligned}
    0 &= ( \partial_- v_+ ) \cdot \Bigg( ( n - 1 ) G + ( X_0 - X_1^2 ) \left( X_1 \frac{\partial F}{\partial X_0} - \frac{\partial G}{\partial X_0} \right) 
    \\
    &\qquad \qquad \qquad \qquad + \nu E'' \left( n F - X_1 \frac{\partial F}{\partial X_1} + \frac{\partial G}{\partial X_1} - X_0 \frac{\partial F}{\partial X_0} + \frac{X_0}{X_1} \frac{\partial G}{\partial X_0} \right) \Bigg) 
    \\
    &\quad + ( \partial_+ v_- ) \cdot \Bigg( ( 1 - n ) X_1 G + ( X_0 - X_1^2 ) \left( X_1 \frac{\partial G}{\partial X_0} - \frac{\partial F}{\partial X_0} \right) 
    \\
    &\qquad \qquad \qquad \qquad + \nu E'' \left( \frac{\partial F}{\partial X_1} - ( n - 2 ) G - X_1 \frac{\partial G}{\partial X_1} + \frac{X_0}{X_1} \frac{\partial F}{\partial X_0} - X_0 \frac{\partial G}{\partial X_0} \right) \Bigg) \, .
\end{aligned}
\end{equation}
This gives rise to four partial differential equations for the functions $F$ and $G$, since for each of the two quantities multiplying $\partial_\pm v_\mp$, the terms proportional to $E''$ must vanish independently. Simultaneously solving the first and third of these equations,
\begin{equation}
\begin{aligned}
    ( n - 1 ) G + ( X_0 - X_1^2 ) \left( X_1 \frac{\partial F}{\partial X_0} - \frac{\partial G}{\partial X_0} \right) &= 0 \, ,
    \\
    ( 1 - n ) X_1 G + ( X_0 - X_1^2 ) \left( X_1 \frac{\partial G}{\partial X_0} - \frac{\partial F}{\partial X_0} \right) &= 0 \, ,
\end{aligned}
\end{equation}
gives the constraints
\begin{align}\label{thm_first_solns}
    F ( X_0, X_1 ) = F ( X_1 ) \, , \qquad G ( X_0, X_1 ) = ( X_0 - X_1^2 )^{n-1} \widetilde{G} ( X_1 ) \, .
\end{align}
Using (\ref{thm_first_solns}), the remaining two differential equations arising from (\ref{four_pde_eqn_thm}) become
\begin{align}\label{pdes_consistency}
    &X_1 F'(X_1) \!-\! n F ( X_1 ) \!=\! \frac{1}{X_1} \left( X_0 \!-\! X_1^2 \right)^{n-2} \left( ( n \!-\! 1 ) ( X_0 \!-\! 2 X_1^2 ) \widetilde{G} ( X_1 ) \!+\! X_1 ( X_0 \!-\! X_1^2 ) \widetilde{G}' ( X_1 ) \right) \, , 
    \notag \\
    \\
    &F'(X_1) \!=\! - ( X_0 \!-\! X_1^2 )^{n-2} \left( \widetilde{G} ( X_1 ) \cdot \left( ( 3 \!-\! 2 n ) X_0 \!+\! ( 3n \!-\! 4 ) X_1^2 \right) \!-\! X_1 ( X_0 \!-\! X_1^2 ) \widetilde{G}' ( X_1 ) \right) \, .
    \notag
\end{align}
The left sides of both lines of equation (\ref{pdes_consistency}) depend only on $X_1$, while the right sides depend on both $X_0$ and $X_1$. Thus, for generic $n$, these differential equations are inconsistent and admit no solution besides the trivial one $F = G = 0$. This concludes the proof that no ansatz for the higher-spin currents involving only the interaction function and finitely many of its derivatives obeys the conservation equation for $n \geq 3$.
\end{proof}

Let us comment on the case $n = 2$ in the proof of Theorem \ref{nogo_thm}. For this value, the prefactors $\left( X_0 \!-\! X_1^2 \right)^{n-2}$ appearing on the right sides of both lines of equation (\ref{pdes_consistency}) are equal to $1$. The remaining equations admit the solution
\begin{align}
    F ( X_0, X_1 ) = - 1 + X_1^2 \, , \qquad G ( X_0, X_1 ) = 2 \left( X_1 - \frac{X_0}{X_1} \right) \, ,
\end{align}
which is the energy-momentum tensor (\ref{stress_tensor_boson}). Thus the stress tensor, but none of the higher-spin currents, can be obtained from a simple ansatz involving only the interaction function and finitely many of its derivatives.

The negative conclusion of Theorem \ref{nogo_thm} is interesting, in part, because it establishes that auxiliary field sigma models do \emph{not} exhibit the same structure as the theories studied in \cite{Lacroix:2017isl}. In that work it was shown that, for a large class of integrable sigma models with Lax connection with spatial component $\mathfrak{L}_\sigma (z ) $ and twist function $\varphi ( z )$, local higher-spin conserved currents can be extracted by evaluating the quantities
\begin{align}\label{sylvain_quantity}
    \mathscr{J}_n = \tr \left( \varphi ( z )^n \mathfrak{L}_\sigma ( z )^n \right)
\end{align}
at the poles of the Lax connection. For auxiliary field deformations of the principal chiral model, the Lax connection $\mathfrak{L}_\pm$ is written in terms of only two structures, the Maurer-Cartan form $j_\pm$ and the quantity $\mathfrak{J}_\pm = - ( j_\pm + 2 v_\pm )$. Using the equations of motion, these two spin-$1$ objects can be expressed entirely in terms of $v_\pm$ and the first derivative $E'$ of the interaction function. Therefore, if higher-spin conserved currents could be written in terms of quantities like (\ref{sylvain_quantity}), they would necessarily be expressible in the form of the ansatz (\ref{theorem_ansatz}) with $N = 1$. However, we have just seen that this is not the case, even for the simplest scenario of auxiliary field deformations of a free boson. Thus, despite the fact that AF sigma models are examples of an $r/s$ system with twist function -- and, in fact, they have the \emph{same} twist functions as the corresponding seed theories -- the analysis of \cite{Lacroix:2017isl} does not apply to them, and one must use other techniques to extract local higher-spin currents.

\subsection{Smirnov-Zamolodchikov flows}\label{subsec:perturbative-analysis-free-boson}

In this section, we turn to our second goal in the case of the single free boson. More precisely, we will analyse the system \eqref{system-original-form}, resulting from the current conservation condition \eqref{conservation}, and the Smirnov-Zamolodchikov flow equation \eqref{flow} at the same time, introducing a constraint on the interaction function $E$ inspired by the construction of leading-order deformations by higher-spin currents discussed in \cite{Bielli:2024ach,Bielli:2024fnp,Bielli:2024oif}. As reviewed around equation \eqref{E-leading-order}, in these works it was shown that by choosing interaction functions of the form
\begin{equation}\label{E-desired-behaviour}
E(\nu_{2},...,\nu_{N}) = \lambda_{n} \nu_{n} 
\end{equation}
one can obtain deformations of the seed theory by spin-$n$ conserved currents, to first order in the deformation parameter $\lambda_{n}$. It is then natural, in connection with this argument, to start looking for solutions of the conservation and flow equations, for which the interaction function is analytic in the deformation parameter and exhibits the leading order behaviour in \eqref{E-desired-behaviour}. For the AF-boson, where there is a single independent variable $\nu$ and one has the identification $\nu_{n}=\nu^{n}$, the dimensions
\begin{equation}
[E]=2
\qquad , \qquad
[\nu]=2
\qquad , \qquad 
[\lambda_{n}]=2-2n \, ,
\end{equation}
naturally bring us to consider interaction functions of the form
\begin{equation}
E(\nu) = \lambda_{n}\nu^{n} \epsilon(\lambda_{n}\nu^{n-1}) \, .
\end{equation}
In this expression $\epsilon$ is an unknown analytic function of the dimensionless variable $z\!:=\!\lambda_{n}\nu^{n-1}$, which at leading order in $\lambda_{n}$ must recover \eqref{E-desired-behaviour}, and the correct dimensionality of $E$ is ensured by the prefactor, which fully encodes the desired leading order behaviour.
In turn, the above reasoning leads us to consider the following expressions for $f$ and $g$
\begin{equation}\label{ansatz-1}
f(\nu) = \phi(\lambda_{n}\nu^{n-1}) \qquad , \qquad g(\nu) = \lambda_{n}\nu^{n} \tilde{\gamma}(\lambda_{n}\nu^{n-1}) \, ,
\end{equation}
where $\phi,\tilde{\gamma}$ are analytic functions and the correct dimensionality of $f,g$ is again ensured by the prefactors. To simplify the next steps, it is convenient at this point to make a redefinition of the function $g(\nu)$, so as to make it dimensionless like $f(\nu)$ and have them on equal footing. Letting $g(\nu) = \nu h(\nu)$, the equations \eqref{system-original-form} and the flow \eqref{flow} become
\begin{equation}\label{conservation+flow-form-used-for-power-series}
\begin{aligned}
\nu h'+(n-1)h &=\nu E'f'-n \nu fE'' \, , 
\\
\nu f'& =E'(\nu h'+(n-1)h)+(n-2)\nu h E'' \, , 
\\
\partial_{\lambda_{n}}E &= \nu^{n}(f^2-h^2) \, ,
\end{aligned}
\end{equation}
the ansatz for the three functions being
\begin{equation}
E(\nu) = \lambda_{n}\nu^{n} \epsilon(\lambda_{n}\nu^{n-1})
\qquad , \qquad 
f(\nu) = \phi(\lambda_{n}\nu^{n-1}) \qquad , \qquad 
h(\nu)=\gamma(\lambda_{n}\nu^{n-1}) \, ,
\end{equation}
with $\epsilon,\phi,\gamma$ dimensionless analytic functions of $z\!:=\!\lambda_{n}\nu^{n-1}$ to be determined and the relation $\gamma(z)=z\tilde{\gamma}(z)$ to \eqref{ansatz-1}. Notice that such rescaling is equivalent to a rewriting of \eqref{free-bosono-higher-spin-currents-guess} as
\begin{equation}
\tau_{\pm n} := f(\nu) v_{\pm}^{n}
\qquad ,\qquad 
\theta_{\pm(n-2)} := h(\nu) v_{\pm}^{n-1}v_{\mp} \, ,
\end{equation}
which is reminiscent of the ansatz used in \cite{Rosenhaus:2019utc}, but in terms of auxiliary fields.

Armed with these new requirements, we are now in a position which easily allows to study the equations \eqref{conservation+flow-form-used-for-power-series} in a perturbative fashion, by considering series expansions of the sought analytic functions.
We thus begin our analysis by considering
\begin{equation}
\epsilon(z)=\sum_{a=0}^{\infty} \epsilon_{a}z^{a}
\qquad , \qquad
\phi(z)=\sum_{a=0}^{\infty} \phi_{a}z^{a}
\qquad , \qquad
\gamma(z)=\sum_{a=0}^{\infty} \gamma_{a}z^{a} \, ,
\end{equation}
and $\epsilon_{a},\phi_{a},\gamma_{a}$ constant coefficients to be determined. This leads to the following rewriting
\begin{equation}\label{functions-desired}
E = \nu\sum_{a=0}^{\infty}\epsilon_{a}z^{a+1}
\qquad , \qquad
f = \sum_{a=0}^{\infty}\phi_{a}z^a
\qquad , \qquad
h = \sum_{a=0}^{\infty}\gamma_{a}z^{a} \, ,
\end{equation}
and, in turn, the relations 
\begin{equation}
\frac{\partial z}{\partial\nu}=(n-1)\nu^{-1}z \, , 
\qquad \qquad \qquad \frac{\partial z}{\partial \lambda_{n}}=\nu^{n-1} \, , 
\end{equation}
allow to expand the derivatives as
\begin{equation}\label{derivatives}
\begin{aligned}
\frac{\mathrm{d}E}{\mathrm{d}\nu}& \!=\! \sum_{a=0}^{\infty}[1\!+\!(n\!-\!1)(a\!+\!1)]\epsilon_{a}z^{a+1} \, ,
\quad \qquad \qquad 
\frac{\mathrm{d}x}{\mathrm{d}\nu} \!=\!\nu^{-1}\sum_{a=0}^{\infty}(n\!-\!1)(a\!+\!1)\chi_{a+1}z^{a+1} \, , 
\\
\frac{\mathrm{d}^{2}E}{\mathrm{d}\nu^2} &\!=\! \nu^{-1}\sum_{a=0}^{\infty}(n\!-\!1)(a\!+\!1)[1\!+\!(n\!-\!1)(a\!+\!1)]\epsilon_{a}z^{a+1} \, , \quad \quad 
\frac{\partial E}{\partial \lambda_{n}} \!=\! \nu^{n} \sum_{a=0}^{\infty} (a\!+\!1)\epsilon_{a}z^{a} \, ,
\end{aligned}
\end{equation}
where $x$ stands for $f,h$ and $\chi$ for $\phi,\gamma$.
At this point the dependence on $\nu$ disappears from \eqref{conservation+flow-form-used-for-power-series} and one is left with expansions in $z$, which can be cast in the form
\begin{equation}\label{eqs-series-form}
\begin{aligned}
\sum_{a=0}^{\infty}(a+1)\gamma_{a}z^{a} & = \sum_{a,b=0}^{\infty} [1+(n-1)(a+1)][b-n(a+1)]\epsilon_{a}\phi_{b}z^{a+b+1} \, ,
\\
\sum_{a=0}^{\infty}(a+1)\phi_{a+1}z^{a+1} & = \sum_{a,b=0}^{\infty} [1+(n-1)(a+1)][(b+1)+(n-2)(a+1)]\epsilon_{a}\gamma_{b}z^{a+b+1} \, ,
\\
\sum_{a=0}^{\infty}(a+1)\epsilon_{a}z^{a} & = \sum_{a,b=0}^{\infty}(\phi_{a}\phi_{b}-\gamma_{a}\gamma_{b})z^{a+b} \, .
\end{aligned}
\end{equation}
Looking at the first two equations, which correspond to \eqref{system-original-form}, it is now even more evident that a solution for $\{\gamma_{a},\phi_{a}\}$ should exist, at every order in $z$, for any choice of $\{\epsilon_{a}\}$. We will now determine such solution in terms of two recursive relations for $\gamma_{a}$ and $\phi_{a}$, only then making use of the flow equation, which plays the role of a constraint on  the resulting coefficients.
From the first equation we immediately recognise that $O(z^0)$ terms lead to the condition $\gamma_{0}=0$, which substituted in the $O(z)$ terms of the second equation leads to $\phi_{1}=0$. Exploiting then the following rewriting
\begin{equation}
\sum_{a,b=0}^{\infty} \Bigg{|}_{a+b=k} x_{a}y_{b} = \sum_{a=0}^{k}x_{a}y_{k-a} = \sum_{b=0}^{k}y_{b}x_{k-a} \, ,
\end{equation}
for $x_{a}$ and $y_{b}$ generic objects depending on $a$ and $b$, the above equations can be rewritten as order-by-order conditions on the coefficients on the left hand side. At any $O(z^k)$ we have
\begin{equation}\label{coefficients-order-k}
\begin{aligned}
(k+1)\gamma_{k}&=-nk\phi_{0}\tilde{\epsilon}_{k-1}+\sum_{b=1}^{k-1}\Bigl[1-\tfrac{n(k-b)}{b}\Bigr] b\phi_{b}\tilde{\epsilon}_{k-1-b}
\qquad \forall \, k\geq 1
\,,
\\
k \phi_{k} & = \sum_{b=1}^{k-1}\Bigl[1+\tfrac{(n-2)(k-b)}{(b+1)}\Bigr](b+1)\gamma_{b} \tilde{\epsilon}_{k-1-b} \qquad \quad\quad \forall \, k\geq 1
\,,\\
(k+1)\epsilon_{k} &= \sum_{b=0}^{k}(\phi_{b}\phi_{k-b}-\gamma_{b}\gamma_{k-b}) \qquad\qquad \qquad \qquad \quad \,\,\,\,  \forall \, k\geq 0 \, ,
\end{aligned}
\end{equation}
where we introduced the shorthand notation $\tilde{\epsilon}_{a}:=[1+(n-1)(a+1)]\epsilon_{a}$ and in the first two equations explicitly separated the $\phi_{0},\gamma_{0}$ contributions from \eqref{eqs-series-form}, using then $\gamma_{0}=0$. Notice that the two summations $\sum_{b=1}^{k-1}$ are understood to vanish for $k=1$, hence reproducing $\phi_{1}=0$ and $\gamma_{1}\propto \phi_{0}$. In this form it is clear that one can substitute the first condition into the second to obtain a recursion relation for $\phi_{k}$ and, vice versa, the second condition into the first to obtain a recursion relation for $\gamma_{k}$. The outcome reads
\begin{align}\label{recursions}
(\!k\!+\!1\!)\gamma_{k} \!&=\! -nk\phi_{0}\tilde{\epsilon}_{k-1} \!+\!\!\sum_{b=1}^{k-1}\!\sum_{c=1}^{b-1}\!\Bigl[\!1\!-\!\tfrac{n(k-b)}{b}\!\Bigr]\!\Bigl[\!1\!+\!\tfrac{(n-2)(b-c)}{(c+1)}\!\Bigr](c\!+\!1)\gamma_{c}\tilde{\epsilon}_{k-1-b}\tilde{\epsilon}_{b-1-c} \, , 
\\
k\phi_{k} \!&=\! -n\phi_{0}\!\!\sum_{b=1}^{k-1}b\Bigl[\!1\!+\!\tfrac{(n-2)(k-b)}{(b+1)}\!\Bigr]\tilde{\epsilon}_{k-1-b}\tilde{\epsilon}_{b-1} \!+\!\! \sum_{b=1}^{k-1}\!\sum_{c=1}^{b-1}\!\Bigl[\!1\!+\! \tfrac{(n-2)(k-b)}{(b+1)}\!\Bigr]\!\Bigl[\!1\!-\!\tfrac{n(b-c)}{c}\!\Bigr]c\phi_{c}\tilde{\epsilon}_{k-1-b}\tilde{\epsilon}_{b-1-c} .
\notag
\end{align}
We stress again that these two recursions are now decoupled and allow to separately determine $\gamma_{k},\phi_{k}$ at any order $k$ purely in terms of the respective lower order coefficients, for any set $\{\epsilon_{a}\}$. If no restrictions are imposed on the latter, the final result is generically very complicated and non-vanishing at any order $k$. This is in agreement with the fact that the two ODEs \eqref{system-original-form}, obtained by imposing conservation of the higher-spin currents, should admit a solution for any choice of interaction function $E$. Imposing compatibility of the currents with the flow, namely including the third equation in \eqref{eqs-series-form} in the analysis, highly constraints the whole system, finally leading to the following restrictions
\begin{equation}\label{constraints-coefficients-single-boson}
\gamma_{2k}=0\,, \qquad \qquad \phi_{2k+1} =0\,, \qquad \qquad \epsilon_{2k+1}=0\,, \qquad \qquad \forall \, k \in \mathbb{N} \, ,
\end{equation}
and determining all the remaining coefficients as complicated functions of $n$, the spin of the sought currents, and the single independent component $\phi_{0}$. The latter is also required, from the flow equation, to satisfy $\epsilon_{0}=\phi_{0}^2$: since we explicitly demanded our interaction function to behave as \eqref{E-desired-behaviour} for $\lambda_{n}\rightarrow 0$, we can always reabsorb $\epsilon_{0}$ in a redefinition of the deformation parameter, which corresponds to setting $\phi_{0}=\pm1$. Using computer algebra, e.g. Mathematica, it is straightforward to compute the coefficients $\gamma_{k},\phi_{k},\epsilon_{k}$ to any desired order $k$, but resumming the series and obtaining closed form expressions for $f,h,E$ in \eqref{functions-desired} seems a very complicated task -- consistently with the difficutly encountered in solving the original system \eqref{system-original-form}. We report below the first few coefficients for generic $\phi_{0}$
\begin{equation}
\begin{aligned}
\gamma_{1}= -\frac{1}{2}n^2\phi_{0}^3 \,,
\qquad & \qquad 
\gamma_{3}=\frac{1}{4}n^5\phi_{0}^7[7n-5]
\,,
\\
\phi_{2}=-\frac{1}{4}n^4\phi_{0}^5
\,,
\qquad & \qquad \phi_{4}=\frac{1}{32}n^6\phi_{0}^9[n(41n-36)+4]
\,,
\\
\epsilon_{2}=-\frac{3}{4}n^4\phi_{0}^6\,,
\qquad & \qquad \epsilon_{4}=\frac{1}{8}n^6\phi_{0}^{10}[7n(5n-4)+2] \, .
\end{aligned}
\end{equation}

\subsubsection*{\ul{\it Integrating out the auxiliary fields}}

To finally obtain the deformed Lagrangian which solves the flow equation \eqref{flow}, one needs to integrate out the auxiliary fields from \eqref{Lagrangian-free-boson}, which we repeat here for simplicity
\begin{equation}
\mathcal{L}_{\varphi}^{\text{E}} = \frac{1}{2}\partial_{+}\varphi \partial_{-} \varphi+v_{+}v_{-}+v_{+}\partial_{-}\varphi + \partial_{+}\varphi v_{-} + E(\nu) \qquad \text{with} \qquad \nu:=v_{+}v_{-} \, .
\end{equation}
The EOM for the auxiliary fields \eqref{EOM} immediately imply the relations
\begin{equation}\label{boson-rel}
\partial_{+}\varphi v_{-}\deq-\nu(1+E')\,,
\qquad \quad
v_{+}\partial_{-}\varphi \deq-\nu(1+E')\,,
\qquad \quad
\partial_{+}\varphi \partial_{-}\varphi\deq\nu(1+E')^2 \,,
\end{equation}
which substituted in the Lagrangian lead to
\begin{equation}\label{boson-L-intermediate}
\mathcal{L}^{E}_{\varphi}\Bigl{|}_{\delta v_{\pm}=0} \deq -\tfrac{1}{2}\nu \Bigl(1-(E')^2+2E'-2\nu^{-1}E\Bigr) \, .
\end{equation}
At this point, using the expansions \eqref{functions-desired} and \eqref{derivatives} one obtains
\begin{equation}\label{lagrangian-sum}
\mathcal{L}^{E}_{\varphi}\Biggl{|}_{\delta v_{\pm}=0} \!\!\!\!\!\!\!\!\!\deq\!  -\tfrac{1}{2}\nu \Bigl(\!1 \!-\!\! \sum_{a,b=0}^{\infty}\!\Bigl[\!1 \!+\! (n\!-\!1)(a\!+\!1)\!\Bigr]\!\Bigl[\!1\!+\!(n\!-\!1)(b\!+\!1)\!\Bigr]\epsilon_{a}\epsilon_{b}z^{a+b+2}+2\!\!\sum_{a=0}^{\infty}(n\!-\!1)(a\!+\!1)\epsilon_{a}z^{a+1}\!\Bigr) \, .
\end{equation}
In this expression one should recall the definition $z\!:=\!\lambda_{n}\nu^{n-1}$ and notice that the variable $\nu:=v_{+}v_{-}$ appears at all orders of the series and still remains undetermined. To obtain the final expression for the Lagrangian one thus needs to compute $\nu$: this can be done by means of another recursion, since the EOM for the auxiliary fields \eqref{EOM} can be written as
\begin{equation}
\begin{aligned}
v_{\pm} & \deq -\partial_{\pm}\varphi -E'v_{\pm} 
\,,
\\
& \deq -\partial_{\pm}\varphi-\lambda_{n}v_{\pm}^{n}v_{\mp}^{n-1}\sum_{a=0}^{\infty}\Bigl[1+(n-1)(a+1) \Bigr]\epsilon_{a}\lambda_{n}^{a}v_{+}^{a(n-1)}v_{-}^{a(n-1)} \,,
\end{aligned}
\end{equation}
where in the second line we have made explicit the expansion \eqref{derivatives} for $E'$ and exploited the definition of $z$. Given the latter expression, one can then easily compute $v_{\pm}$ to any desired order in $\lambda_{n}$ by recursively substituting the expression into itself. With the help of computer algebra, e.g. Mathematica, one can immediately obtain the first few orders
\begin{equation}
\begin{aligned}
v_{\pm}\deq& -\partial_{\pm}\varphi+n(\partial_{\pm}\varphi)^{n}(\partial_{\mp}\varphi)^{n-1}\lambda_{n}-n^2(2n-1)(\partial_{\pm}\varphi)^{2n-1}(\partial_{\mp}\varphi)^{2n-2} \lambda_{n}^2+
\\
& +\tfrac{n^3}{4}(3n-2)(7n-4)(\partial_{\pm}\varphi)^{3n-2}(\partial_{\mp}\varphi)^{3n-3}\lambda_{n}^3 + 
\\
& + \tfrac{n^4}{6}(4n-3)(10-30n+23n^2)(\partial_{\pm}\varphi)^{4n-3}(\partial_{\mp}\varphi)^{4n-4}\lambda_{n}^4 +O(\lambda_{n}^5) \, .
\end{aligned}
\end{equation}
In turn, this allows to compute $\nu:=v_{+}v_{-}$ and substitute back in \eqref{lagrangian-sum}, where the coefficients $\{\epsilon_{a}\}$ are given by the recursion \eqref{recursions} combined with the third relation in \eqref{coefficients-order-k}. The Lagrangian then reads, fixing $\epsilon_{0}=\phi_{0}=1$ for convenience,
\begin{align}\label{O(4)-boson-lagrangian}
\mathcal{L}^{E}_{\varphi}\Bigl{|}_{\delta v_{\pm}=0} \deq&  -\tfrac{1}{2} \partial_{+}\varphi\partial_{-}\varphi+(\partial_{+}\varphi\partial_{-}\varphi)^{n}\lambda_{n}-n^2 (\partial_{+}\varphi\partial_{-}\varphi)^{2n-1}\lambda_{n}^2
\\
& +\tfrac{n^3}{4}(7n-4)(\partial_{+}\varphi\partial_{-}\varphi)^{3n-2}\lambda_{n}^3 -\tfrac{n^4}{6}(10-30n+23n^2)(\partial_{+}\varphi\partial_{-}\varphi)^{4n-3}\lambda_{n}^4 + O(\lambda_{n}^5) \, .
\notag 
\end{align}
Multiplying \eqref{O(4)-boson-lagrangian} by an overall $-4$ factor and replacing $n\rightarrow s+1$ in each term, the above expression correctly agrees with the one previously found in \cite{Rosenhaus:2019utc}\footnote{An equivalent way to obtain \eqref{O(4)-boson-lagrangian} is exploiting the third relation in \eqref{boson-rel} to rewrite \eqref{boson-L-intermediate} as 
\begin{equation}
\mathcal{L}^{E}_{\varphi}\Bigl{|}_{\delta v_{\pm}=0} \deq -\tfrac{1}{2} (\partial_{+}\varphi\partial_{-}\varphi) + \tfrac{(E')^2}{(1+E')^2}(\partial_{+}\varphi\partial_{-}\varphi) + E \, ,
\end{equation}
then recursively substituting $\nu\deq (\partial_{+}\varphi\partial_{-}\varphi)(1+E')^{-2}$ in $E(\nu),E'(\nu)$ and expanding at the desired order.
}.

This completes the discussion of the second goal of this article, constructing solutions to Smirnov-Zamolodchikov flows, in the special case of deformations of a single free boson. Again, we stress that -- while higher-spin conserved currents exist for \emph{any} choice of interaction function $E$, as established in Section \ref{subsec:free_boson_currents} -- there exists only a countably infinite set of interaction functions $E = E ( \lambda_n , \nu )$ which solve SZ flow equations (\ref{flow}) driven by spin-$n$ combinations and reducing to the free boson when $\lambda_n = 0$. The additional assumption that the interaction function obey such a flow equation leads to the third constraint in (\ref{conservation+flow-form-used-for-power-series}), whereas in the construction of higher-spin currents in Section \ref{subsec:free_boson_currents}, the function $E$ was unconstrained. We have described a recursive procedure to generate both the interaction function $E ( \lambda_n, \nu )$ and the associated Lagrangian $\mathcal{L}$, after integrating out auxiliary fields, for every such interaction function $E$ obeying a spin-$n$ Smirnov-Zamolodchikov flow, which solves the second problem outlined in the introduction for this case.

\section{Spin-$2k$ flows of AF sigma models as free boson flows}\label{sec:spin-2k-of-AFSM}

In this section we take a step back from the explicit toy model previously considered, and exploit general features of AF sigma models to derive an important identity.
This allows us to construct an infinite family of even higher-spin conserved currents for various infinite families of AF sigma models whose interaction functions $E$ depend only on $\nu_2$.\footnote{As pointed out in \cite{Ferko:2024ali}, auxiliary field models whose interaction functions depend only on $\nu_2$ include all deformations of the seed theory by arbitrary functions of the energy-momentum tensor.} We therefore accomplish the first goal of this paper, construction of spin-$n$ conserved currents, for any even $n$ and for any interaction function of one variable in the class of theories.

Crucially, this is achieved by noting that the conservation equation for all such currents can be mapped to the system of coupled 1st order ODEs \eqref{system-original-form} encountered in the case of the AF free boson. Such a connection immediately implies that one can re-use the whole perturbative analysis of Section \ref{subsec:perturbative-analysis-free-boson} to characterize 
tuples $(E, \tau_{\pm 2n} , \theta_{\pm ( 2n - 2 )} )$ of data which satisfy SZ flows (\ref{flow}), accomplishing the second goal of this work for this family of models.

\subsection{Unification of auxiliary field models}\label{sec:unification}

First we will point out a common underlying structure which is shared by many of the examples of auxiliary field sigma models that have been constructed in the literature. This will allow us to prove an important identity (\ref{important-identity}) which holds for all models which exhibit this shared structure. In later subsections, we will see that this identity will be very useful in pursuing both of the main goals of this paper in a much broader class of examples than the single free boson considered in the preceding section.

Consider an AF sigma model $S^{\text{E}}[\phi,v]$ where, in line with previous results, the background $\mathcal{M}$ is constructed out of some Lie group G with Lie algebra $\mathfrak{g}$, such that the fundamental fields are understood as coordinates on $\mathcal{M}$ and the auxiliary fields $v$ are Lie-algebra-valued quantities. If the EOM for the model can be written in the form 
\begin{equation}\label{AF-general-EOM-for-identity}
\begin{aligned}
D_{+}\mathcal{B}_{-}\!+\!D_{-}\mathcal{B}_{+} \deq & \, 0 
\quad\,\,\,\, \text{with} \quad\,\,\,\,
\mathcal{B}_{\pm} := -(\mathcal{A}_{\pm}+2v_{\pm}) \, ,
\\
\mathcal{A}_{\pm}+v_{\pm}+\Delta_{\pm} \deq & \, 0 
\quad\,\,\,\, \text{with} \quad\,\,\,\, 
\Delta_{\pm}:= \delta_{v_{\mp}}E(v) 
\quad \text{and} \quad [\Delta_{\pm},v_{\mp}]\deq \, 0 \, ,
\\
D_{+}\mathcal{A}_{-} \!-\!D_{-}\mathcal{A}_{+} \!+\! a \,[\mathcal{A}_{+},\mathcal{A}_{-}] \deq & \, 0 
\quad\,\,\,\, \text{with} \quad\,\,\,\, 
 D_{\pm}:=\partial_{\pm}\!+\![\mathcal{C}_{\pm},-]
\quad \text{and} \quad a \! \in \! \mathbb{R} \,\, \text{const.} \, , 
\end{aligned}
\end{equation}
for some Lie-algebra-valued one-forms $\mathcal{A},\mathcal{B},\mathcal{C}$,
the following identity holds
\begin{equation}\label{important-identity}
\partial_{\mp} \, \mathrm{tr}(v_{\pm}^{n}) \deq \, n \,\mathrm{tr} (v_{\pm}^{n-1} \partial_{\pm}\Delta_{\mp}) \, .
\end{equation}
To check this, it is sufficient to notice that the second line in \eqref{AF-general-EOM-for-identity} implies the rewriting
\begin{equation}\label{v-Delta}
v_{\pm} \deq -\tfrac{1}{2}(\mathcal{A}_{\pm}+\mathcal{B}_{\pm}) 
\,,\qquad \qquad \Delta_{\pm}\deq -\tfrac{1}{2}(\mathcal{A}_{\pm}-\mathcal{B}_{\pm}) \, ,
\end{equation}
while the first and third line in \eqref{AF-general-EOM-for-identity} imply
\begin{equation}\label{del-(A+B)}
\partial_{\mp}(\mathcal{A}_{\pm}+\mathcal{B}_{\pm})=\partial_{\pm}(\mathcal{A}_{\mp}-\mathcal{B}_{\mp})\pm a \, [\mathcal{A}_{+},\mathcal{A}_{-}]+2[\mathcal{C}_{\mp},v_{\pm}]-2[\mathcal{C}_{\pm},\Delta_{\mp}]  \, . 
\end{equation}
Then using \eqref{v-Delta} and \eqref{del-(A+B)} one finds that
\begin{equation}\label{identity-steps}
\begin{aligned}
\partial_{\mp} \, \mathrm{tr}(v_{\pm}^{n}) \deq &  -\frac{n}{2} \, \mathrm{tr}\Bigl( v_{\pm}^{n-1} \partial_{\mp}(\mathcal{A}_{\pm}+\mathcal{B}_{\pm}) \Bigr)
\\
\deq & -\frac{n}{2} \, \mathrm{tr}\Bigl( v_{\pm}^{n-1} \partial_{\pm}(\mathcal{A}_{\mp}-\mathcal{B}_{\mp}) \Bigr)
\\ 
& \mp \frac{an}{2} \,  \mathrm{tr}\Bigl( v_{\pm}^{n-1} [\mathcal{A}_{+},\mathcal{A}_{-}] \Bigr) -n \, \mathrm{tr}\Bigl(v_{\pm}^{n-1}[\mathcal{C}_{\mp},v_{\pm}]\Bigr) + n \, \mathrm{tr}\Bigl(v_{\pm}^{n-1}[\mathcal{C}_{\pm},\Delta_{\mp}]\Bigr)
\\
\deq & + n \, \mathrm{tr}\Bigl( v_{\pm}^{n-1} \partial_{\pm}\Delta_{\mp} \Bigr) \, ,
\end{aligned}
\end{equation}
which indeed corresponds to the desired identity \eqref{important-identity} after noting, in the last step, the vanishing of the three commutator terms. The one involving $[\mathcal{C}_{\mp},v_{\pm}]$ clearly vanishes by construction, while the other two are explicitly shown to vanish in appendix \ref{appendix:identity}, given \eqref{AF-general-EOM-for-identity}.
One can then notice that from the results in \cite{Ferko:2024ali,Bielli:2024khq,Bielli:2024ach,Bielli:2024fnp,Bielli:2024oif} the conditions \eqref{AF-general-EOM-for-identity} are met for
\begin{itemize}
\item AF PCMs, with and without Wess-Zumino term, using the identification $a=1$ and
\begin{equation}
\mathcal{A}_{\pm}:= j_{\pm} 
\qquad , \qquad 
\mathcal{C}_{\pm}:=0
\qquad \text{with} \qquad 
j_{\pm}:=g^{-1}\partial_{\pm}g \, .
\end{equation}
\item AF T-dual models, using the identification $a=1$ and
\begin{equation}
\mathcal{A}_{\pm} := \tilde{j}_{\pm}
\qquad , \qquad \mathcal{C}_{\pm}:=0
\qquad \text{with} \qquad 
\tilde{j}_{\pm}:= {\pm}\frac{1}{1\pm\mathrm{ad}_{X}}(\partial_{\pm}X\mp 2v_{\pm}) \, .
\end{equation}
\item AF (bi)-Yang-Baxter models, using the identification $a=(1-c^2\eta^2+\tilde{c}^2\zeta^2)$ and
\begin{equation}
\begin{aligned}
\mathcal{A}_{\pm}:=  J_{\pm}^{\zeta}
\qquad &, \qquad 
\mathcal{C}_{\pm}:= \pm\zeta\tilde{\mathcal{R}}(\mathcal{B}_{\pm})
\qquad \text{with} \qquad 
\\
J_{\pm}^{\zeta}:= -(\mathfrak{J}_{\pm}^{\zeta}\!+\!2v_{\pm})
\qquad \text{a}&\text{nd} \qquad 
\mathfrak{J}_{\pm}^{\zeta}\!:=  \!-\frac{1}{1\!\mp\! \eta \mathcal{R}_{g} \!\mp\! \zeta \tilde{\mathcal{R}}}(j_{\pm}\!+\!2v_{\pm}) \, .
\end{aligned}
\end{equation}
\item AF Symmetric Space models,\footnote{The construction seems more involved for semi-symmetric spaces, where fermionic contributions from $\mathcal{A}$ spoil the identity \eqref{important-identity}. The issue might potentially be resolved by allowing for extra fermionic contributions coming from the auxiliary fields, which were not taken into account in the analysis of \cite{Bielli:2024oif}. Extension to $\mathbb{Z}_{N}$ cosets \cite{Cesaro:2024ipq} would also be interesting.}
using the identification $a=0$ and 
\begin{equation}
\mathcal{A}_{\pm}:=j_{\pm}^{(2)}
\quad , \quad 
\mathcal{C}_{\pm}:=j_{\pm}^{(0)} 
\qquad \text{with} \qquad
v_{\pm} \rightarrow v_{\pm}^{(2)} \, .
\end{equation}
\end{itemize}
We conclude this subsection by highlighting other two interesting common features of the AF sigma models discussed above in the frame of the EOM \eqref{AF-general-EOM-for-identity}. The first is that such models exhibit Lax connections which can be rearranged as
\begin{equation}\label{common-Lax}
\mathfrak{L}_{\pm}=l_{1}^{\pm}(z)\frac{\mathcal{A}_{\pm}\pm z\mathcal{B}_{\pm}}{1\pm z} + l_{2}^{\pm}(z)\Theta(v_{\pm})+l_{3}^{\pm}(z) \mathcal{C}_{\pm}
\qquad \text{with} \qquad
\Theta:\mathfrak{g}\rightarrow\mathfrak{g}\, ,
\end{equation}
and suitable choices of functions $l^\pm_1(z)$, $l^\pm_2(z)$, and $l^\pm_3(z)$.
The second is that all their Lagrangians can be written in the form
\begin{equation}\label{common-Lagrangian}
\mathcal{L}(\phi,v) = \frac{1}{2}\mathrm{tr}\Bigl( (\mathcal{K}_{-}+2v_{-})\mathcal{O}_{-} (\mathcal{K}_{+}+2v_{+})\Bigr) - \mathrm{tr}( v_{-}v_{+}) + E(v) \, ,
\end{equation}
with the EOM for the auxiliary fields, namely the second equation in \eqref{AF-general-EOM-for-identity}, becoming
\begin{equation}\label{common-v-EOM}
v_{\pm}+\mathcal{O}_{\pm}(\mathcal{K}_{\pm})+\mathcal{O}_{\mp}^{-1}\mathcal{O}_{\pm}\Delta_{\pm}\deq 0 \, ,
\end{equation}
where $\mathcal{O}_{\pm}\!:\! \mathfrak{g} \!\rightarrow\!\mathfrak{g}$ are operators on the underlying Lie algebra and $\mathcal{K}_{\pm}$ encode the fundamental fields $\phi$ of the theory.
While \eqref{common-Lax} is simply a rewriting of the Lax connection proposed in previous works in terms of the quantities introduced in \eqref{AF-general-EOM-for-identity}, the Lagrangian \eqref{common-Lagrangian} and the EOM \eqref{common-v-EOM} immediately imply that integrating out the auxiliary fields leads to
\begin{equation}\label{integrated-Lagrangian-step-0}
\mathcal{L}(\phi,v) \deq  -\frac{1}{2} \mathrm{tr} (\mathcal{K}_{+}\mathcal{O}_{-}\mathcal{K}_{-})+\mathrm{tr}(\Delta_{+}\mathcal{O}^{-1}_{+}\mathcal{O}_{-}\Delta_{-})+E(v) \, , 
\end{equation}
after exploiting that, in all cases under consideration, one has
\begin{equation}\label{operators-general-form}
\mathcal{O}_{\pm}:=\frac{1}{1\pm M} 
\quad \text{with} \quad 
M^{T}=-M 
\quad \text{and} \quad 
\mathcal{O}^{-1}_{\pm}:=1\pm M \, ,
\end{equation}
for $M$ some operator on the underlying Lie algebra which can be read off each case below.
Obviously, \eqref{integrated-Lagrangian-step-0} is not the full expression for the deformed Lagrangian, since it still depends on $v$ through the interaction function $E$ and its variations $\Delta_{\pm}$. However, this will turn out to be very useful in both (i) characterizing higher-spin conserved currents, and (ii) giving perturbative expressions for solutions to SZ flows, as will become clear in the next sections.
To summarize, we have unified most of the auxiliary field sigma models considered in \cite{Ferko:2024ali,Bielli:2024ach,Bielli:2024fnp,Bielli:2024oif} using an abstraction which emphasizes the common underlying structure of these theories. The Lagrangians and Lax connections for each of the special cases can be recovered from the general expressions above by making the following identifications:
\begin{itemize}
\item AF PCMs
\begin{equation}
\mathcal{K}_{\pm} := j_{\pm} 
\quad , \quad  \mathcal{O}_{\pm}:=\mathds{1}
\quad , \quad 
l_{1}^{\pm}:=\frac{1}{1\mp z} 
\quad , \quad 
l_{2}^{\pm}:= 0 
\quad , \quad 
l_{3}^{\pm}:= 0 \, .
\end{equation}\item AF T-dual models\footnote{Technically, this choice reproduces the Lagrangian of the T-dual model with an extra overall minus sign and the redefinitions $X\rightarrow -X$, $E\rightarrow -E$, which are however purely conventional.} 
\begin{equation}
\mathcal{K}_{\pm} :=\!\pm \partial_{\pm}X 
\quad , \quad 
\mathcal{O}_{\pm}:=\frac{1}{1\!\pm\!\mathrm{ad}_{X}}
\quad , \quad 
l_{1}^{\pm}:=\frac{1}{1\!\mp\! z} 
\quad , \quad 
l_{2}^{\pm}:= 0 
\quad , \quad 
l_{3}^{\pm}:= 0 \, .
\end{equation}
\item AF (bi)-Yang-Baxter models 
\begin{align}
&\mathcal{K}_{\pm} := j_{\pm} 
\quad , \quad \mathcal{O}_{\pm}:=\frac{1}{1\pm \eta \mathcal{R}_{g} \pm \zeta\tilde{{\mathcal{R}}}}
\quad , \quad 
\Theta:=\zeta \tilde{\mathcal{R}} \, , 
\\
& l_{1}^{\pm}:=\pm \Bigl( \frac{2\zeta\tilde{c}\pm(1-\eta^2c^2+\zeta^2\tilde{c}^2)}{1\mp z}-\zeta \tilde{c}\Bigr)
\quad , \quad 
l_{2}^{\pm}:=\mp\frac{2}{1\pm z}
\quad , \quad 
l_{3}^{\pm}:=-\frac{1\mp z}{1\pm z} \, .
\notag
\end{align}
\item AF Symmetric Space models 
\begin{equation}
\begin{aligned}
&\mathcal{K}_{\pm}:=j_{\pm}^{(2)} 
\quad , \quad \mathcal{O}_{\pm}:= \mathds{1} 
\quad , \quad 
\Theta:= \mathds{1} \, ,
\\
& l_{1}^{\pm}:=-1 
\quad , \quad 
l_{2}^{\pm}:= \pm \frac{2z}{1\mp z} 
\quad , \quad 
l_{3}^{\pm}:=1 \, .
\end{aligned}
\end{equation}
\end{itemize}

\subsection{Even higher-spin currents}\label{sec:even_higher_spin}

We will now discuss the first goal of this paper -- construction of higher-spin conserved currents -- for any auxiliary field sigma model that exhibits the ``unified'' structure described in the last subsection. However, we will restrict our attention to spin-$n$ currents for even integers $n = 2k$, and focus on even currents which take a particular form. In general, there can be many spin-$n$ local conserved currents in an integrable field theory. For instance, in the undeformed PCM, both $\tr ( j_\pm^{2n} )$ and $T_{\pm \pm}^{n}$ are conserved spin-$2n$ currents; in this section we will construct currents with a structure more similar to the latter, while in section \ref{sec:su3} we will see an odd spin example with a trace structure like the former (more specifically deformations of $\ttr{j_\pm^3}$) which makes some intricacies of that case clear. We also restrict the form of the interaction function $E$ so that, rather than depending on all of the quantities $\nu_n$ defined in equation (\ref{nuk_defn}) (or even more general functional forms like (\ref{newE})), it is only a function of the single invariant $\nu_2 := \tr ( v_+^2 ) \tr ( v_-^2 )$. 
For models with an interaction function of this form, one has
\begin{equation}\label{Delta-nu2}
\Delta_{\pm} \deq 2 E'(\nu_{2}) \mathrm{tr}(v_{\pm}^2) v_{\mp} \, .
\end{equation}
The results of the last subsection open up the possibility of constructing a simple ansatz for even higher-spin currents of any AF sigma model in our unified family, i.e. those satisfying \eqref{AF-general-EOM-for-identity}, and with $E = E ( \nu_2 )$. The ansatz for the currents take the form\footnote{To avoid confusion, we emphasize that throughout this paper we adopt a concise notation such as $\mathrm{tr}(v_{\pm}^p)^q$ to denote $\left[\mathrm{tr}(v_{\pm}^p)\right]^q$.}
\begin{equation}\label{ansatz-even-flows}
\tau_{\pm2k}:=f(\nu_{2})\mathrm{tr}(v_{\pm}^2)^k \, , \qquad \qquad \theta_{\pm(2k-2)}:=g(\nu_{2})\mathrm{tr}(v_{\pm}^{2})^{k-1} \, ,
\end{equation}
and one can immediately notice that the identity \eqref{important-identity} can be exploited, for $n=2$, to simplify the conservation equation \eqref{conservation}. Using \eqref{important-identity} and \eqref{Delta-nu2} one indeed finds
\begin{equation}
        \begin{split}
            \del _+\, \mathrm{tr}(v_-^2)&\deq4\left (E''\, \nu _2+ E' \right )\mathrm{tr}(v_-^2)\,\del _- \,\mathrm{tr}(v_+^2)+ 2\left (2\, E''\,\nu _2+E' \right )\, \mathrm{tr}(v_+^2)\, \del _-\, \mathrm{tr}(v_-^2)\,,\\[0.5em]
            \del _-\, \mathrm{tr}(v_+^2)&\deq4\left (E''\, \nu _2+ E' \right )\mathrm{tr}(v_+^2)\,\del _+ \,\mathrm{tr}(v_-^2)+ 2\left (2\, E''\,\nu _2+E' \right )\, \mathrm{tr}(v_-^2)\, \del _+\, \mathrm{tr}(v_+^2) \,,
        \end{split}
\end{equation}
and solving these relations for $\del_- \, \mathrm{tr}(v_-^2)$ and $\del _+ \mathrm{tr}(v_+^2)$ leads to
\begin{equation}\label{derivatives-2k-flows}
    \begin{split}
        \del _- \, \mathrm{tr}(v_-^2)&= \frac 1 {\mathrm{tr}(v_+^2)}\, \frac 1 {2(2\, E'' \, \nu_2+ E')}\left ( \del _+ \, \mathrm{tr}(v_-^2)- 4(E'+E''\, \nu _2 )\mathrm{tr}(v_-^2)\, \del_- \mathrm{tr}(v_+^2)\right ),\\[0.5em]
         \del _+ \, \mathrm{tr}(v_+^2)&= \frac 1 {\mathrm{tr}(v_-^2)}\, \frac 1 {2(2\, E'' \, \nu_2+ E')}\left ( \del _- \, \mathrm{tr}(v_+^2)- 4(E'+E''\, \nu _2 )\mathrm{tr}(v_+^2)\, \del_+ \mathrm{tr}(v_-^2)\right ).
    \end{split}
\end{equation}
One can then compute the derivatives of the two currents
\begin{equation}
    \begin{split}
            \del _- \, \tau _{(2k)}&= \left ( f' \, \nu _2+ k \, f\right )\, \mathrm{tr}(v_+^2)^{k-1}\del _-\, \mathrm{tr}(v_+^2)+ f' \, \mathrm{tr}(v_+^2)^{k+1}\, \del _- \, \mathrm{tr}(v_-^2)\,,\\[0.5em]
            \del _+ \, \theta _{(2k-2)}&=\left ( g' \, \nu _2+ (k-1)\, g\right ) \, \mathrm{tr}(v_+^2)^{k-2} \del _+\, \mathrm{tr}(v_+^2)+ g' \, \mathrm{tr}(v_+^2)^k\,  \del _+\, \mathrm{tr}(v_-^2) \,,
    \end{split}
\end{equation}
so that substituting \eqref{derivatives-2k-flows} and rearranging terms one arrives at
\begin{equation}
\begin{aligned}
& \partial_-{\tau_{+2k}}+\partial_{+}\theta_{+(2k-2)} 
\\
& = \mathrm{tr}(v_{+}^2)^{k-1}\partial_{-} \, \mathrm{tr}(v_{+}^2)\Bigl[ \nu_{2} f' + kf + \frac{\nu_{2} g' + (k-1)g}{\nu_{2}(4 E''\nu_{2}+2E')} - \frac{f'\nu_{2}(4E''\nu_{2}+4E')}{(4 E''\nu_{2}+2E')}\Bigr]+
\\
& +\mathrm{tr}(v_{+}^2)^k\partial_{+} \, \mathrm{tr}(v_{-}^2)\Bigl[ g'- (4E''\nu_{2}+4E') \frac{\nu_{2} g' + (k-1)g}{\nu_{2}(4 E''\nu_{2}+2E')} + \frac{f'}{(4 E''\nu_{2}+2E')}   \Bigr] \, ,
\end{aligned}
\end{equation}
and conservation imposes that
\begin{equation}\label{eqs-even-flows}
\begin{aligned}
\nu_{2} f' + kf + \frac{\nu_{2} g' + (k-1)g}{\nu_{2}(4 E''\nu_{2}+2E')} - \frac{f'\nu_{2}(4E''\nu_{2}+4E')}{(4 E''\nu_{2}+2E')} & =0 \, , 
\\
g'- (4E''\nu_{2}+4E') \frac{\nu_{2} g' + (k-1)g}{\nu_{2}(4 E''\nu_{2}+2E')} + \frac{f'}{(4 E''\nu_{2}+2E')}   & = 0 \, .
\end{aligned}
\end{equation}
At this point, looking at the initial ansatz \eqref{ansatz-even-flows} and the conservation conditions \eqref{eqs-even-flows}, it is not hard to notice a close resemblance with the expressions studied in the context of the free boson, namely \eqref{free-bosono-higher-spin-currents-guess} and \eqref{system-original-form}. Indeed, except for a difference in dependence -- $\nu_{2}:=\mathrm{tr}(v_{+}^2)\mathrm{tr}(v_{-}^2)$ here versus $\nu:=v_{+}v_{-}$ in the case of the boson with no underlying Lie algebra structure -- and in the order of the sought currents -- $2k$ here versus $n$ in the boson -- many pieces in the two set of ODEs look exactly the same. This similarity can in fact be made stronger, by simply noting that renaming $n\equiv 2k$ and $\nu \equiv \sqrt{\nu_{2}}$ in \eqref{system-original-form} leads to
\begin{equation}
\begin{aligned}
\nu f'(\nu) &= 2\nu_{2}f'(\nu_{2}) \, , 
\quad \quad \qquad   
\nu g'(\nu)+(n-2)g(\nu)=2[\nu_{2}g'(\nu_{2})+(k-1)g(\nu_{2})] \, , 
\\
\nu^2E''(\nu)&=\nu_{2}[4\nu_{2}E''(\nu_{2})+2E'(\nu_{2})] \, , 
\quad \quad   \;\; 
(\nu E'(\nu))' = \sqrt{\nu_{2}}[4\nu_{2}E''(\nu_{2})+4E'(\nu_{2})] \, , 
\end{aligned}
\end{equation}
which in turn completely maps the system \eqref{system-original-form} onto \eqref{eqs-even-flows}. In light of this property, we are immediately allowed to reuse all the results obtained in section \ref{sec:free_boson} for the AF free boson. In particular, the existence and uniqueness theorems for the system of ordinary differential equations \eqref{system-original-form} guarantee us that one can always find a solution for the functions $f$ and $g$ which determine each of the local higher-spin conserved currents in this model.

This completes the solution of the first problem of our interest, constructing local spin-$n$ currents, for any auxiliary field sigma model obeying \eqref{AF-general-EOM-for-identity} whose interaction function depends only on $\nu_2 = \tr ( v_+^2 ) \tr ( v_-^2 )$ and for any even $n$.

\subsection{Even Smirnov-Zamolodchikov flows}\label{sec:even-SZ-flows}

Let us now turn to the study of Smirnov-Zamolodchikov flows, the second objective, for the case of interest in this section. As we have just seen, all of the observations of section \ref{sec:free_boson} for the AF single free boson also hold here, including the recursive solution \eqref{recursions} and constraints \eqref{constraints-coefficients-single-boson} which characterize the triplet of data $(E, \tau_{\pm 2k}, \theta_{\pm ( 2 k - 2 )} )$ describing the solution to a Smirnov-Zamolodchikov flow. Remarkably, since this whole construction solely relied on the assumptions \eqref{AF-general-EOM-for-identity}, the results obtained for the free boson can again immediately be reused for all classes of AF sigma models which meet the necessary requirements, just as in the construction of the higher-spin currents in section \ref{sec:even_higher_spin}.

In practical terms, this means that given the functions $E(\nu),f(\nu),g(\nu)$ solving the conservation and flow equation \eqref{conservation+flow-form-used-for-power-series} for the single free boson
\begin{equation}\label{E_f_g_expansion_even_flow}
\begin{aligned}
E(\nu)=&\sum_{a=0}^{\infty} \epsilon_{2a}\nu^{2a(n-1)+n} \lambda_{n}^{2a+1} \, , 
\\
f(\nu)=\sum_{a=0}^{\infty}\phi_{2a}\nu^{2a(n-1)}\lambda_{n}^{2a} \, , 
\qquad & \qquad 
g(\nu)=\sum_{a=0}^{\infty}\gamma_{2a+1}\nu^{2a(n-1)+n}\lambda_{n}^{2a+1} \, ,
\end{aligned}
\end{equation}
with the coefficients given via \eqref{recursions} and the last equation in \eqref{coefficients-order-k}, one can immediately obtain the solution to \eqref{eqs-even-flows} by replacing $n\rightarrow2k$ and $\nu\rightarrow\sqrt{\nu_{2}}$
\begin{equation}\label{E-f-g-spin-2k}
\begin{aligned}
E(\nu_2)=&\sum_{a=0}^{\infty} \epsilon_{2a}\nu_{2}^{a(2k-1)+k} \lambda_{2k}^{2a+1} \, , 
\\
f(\nu_{2})=\sum_{a=0}^{\infty}\phi_{2a}\nu_{2}^{a(2k-1)}\lambda_{2k}^{2a} \, , 
\qquad & \qquad 
g(\nu_{2})=\sum_{a=0}^{\infty}\gamma_{2a+1}\nu_{2}^{a(2k-1)+k}\lambda_{2k}^{2a+1} \,,
\end{aligned}
\end{equation}
solving at the same time the flow equation
\begin{equation}
\frac{\partial E(\nu_2)}{\partial\lambda_{2k}}=\nu_2^k f(\nu_{2})^2-\nu_{2}^{k-1}g(\nu_2)^2 \, .
\end{equation}
Notice that the leading order behaviour of the interaction function in \eqref{E-f-g-spin-2k} is $E(\nu_{2})\simeq \lambda_{2k}\nu_{2}^k$, which is quite different from $E(\nu_{2k})\simeq \lambda_{2k}\nu_{2k}$ that one would have imposed in trying to construct spin-$2k$ flows using the previously established relation to SZ flows in \eqref{E-desired-behaviour}. While the former gives us for free -- thanks to the free boson analysis -- spin-$2k$ SZ flows generated by $k$-th powers of spin-2 currents, the latter would require the construction of possibly more general spin-$2k$ currents and would thus require a dedicated analysis. It should also be stressed that while the original variable $\nu:=v_{+}v_{-}$ has no underlying Lie algebraic structure, after performing the mapping -- which formally consists of a simple renaming -- one ends up with $\nu_{2}:=\mathrm{tr}(v_{+}^2)\mathrm{tr}(v_{-}^2)$, which is considerably more complicated and Lie algebra dependent.  The framework established in section \ref{sec:unification} allows at this point to use the interaction function \eqref{E_f_g_expansion_even_flow} to determine the deformed theory, after integrating out the auxiliary fields from the generic AFSM Lagrangian
\begin{equation}\label{c}
\mathcal{L}(\phi,v) \deq  -\frac{1}{2} \mathrm{tr} (\mathcal{K}_{+}\mathcal{O}_{-}\mathcal{K}_{-})+\mathrm{tr}(\Delta_{+}\mathcal{O}^{-1}_{+}\mathcal{O}_{-}\Delta_{-})+E(\nu_{2}) \, , 
\end{equation}
using the auxiliary field EOM
\begin{equation}\label{a}
v_{\pm}+\mathcal{O}_{\pm}(\mathcal{K}_{\pm})+\mathcal{O}_{\mp}^{^{-1}}\mathcal{O}_{\pm}\Delta_{\pm}\deq 0 \, , 
\end{equation}
and the fact that 
\begin{equation}\label{b}
\Delta_{\pm} \deq 2 E'(\nu_{2}) \mathrm{tr}(v_{\pm}^2) v_{\mp} \, .
\end{equation}
The calculation is in principle much more involved than the free boson case, due to the presence of traces and Lie-algebra-valued structures of different natures, which mix in various ways at every order of the perturbative expansion. It is however not too complicated to realise that there exists a finite and closed set of structures arising at every new order in $\lambda_{2k}$, since the AFSM Lagrangian \eqref{c} only depends on the auxiliary fields via $\nu_{2}:=\mathrm{tr}(v_{+}^2)\mathrm{tr}(v_{-}^2)$ and $\Delta_{\pm}$. One should look for relations involving these two quantities and combining \eqref{b} with \eqref{a} it is simple to rewrite the middle term in \eqref{c} as
\begin{equation}\label{d}
\mathrm{tr}(\Delta_{+}\mathcal{O}^{-1}_{+}\mathcal{O}_{-}\Delta_{-}) \deq -2E'\mathrm{tr}(v_{-}^2)\mathrm{tr}\Bigl( (\mathcal{O}_{-}\mathcal{K}_{+})\Delta_{+} \Bigr)-8(E')^3\nu_{2}^2 \, .
\end{equation}
While the second term on the right hand side of \eqref{d} clearly depends on $\nu_2$ only, the first one also exhibits an explicit dependence on $\mathrm{tr}(v_{-}^2)$ -- one of the building blocks of $\nu_2$ -- and a new structure which combines the fundamental fields in $\mathcal{K}_{\pm}$ with the auxiliaries in $\Delta_{\pm}$. It is then natural to start looking at the building blocks of $\nu_2$, which using \eqref{a} and then \eqref{b} can be written as
\begin{equation}\label{vpm2}
\mathrm{tr}(v_{\pm}^2)\deq \mathrm{tr}\Bigl(\! (\mathcal{O}_{\pm}\mathcal{K}_{\pm})^2 \!\Bigr)+2\mathrm{tr}\Bigl(\!(\mathcal{O}_{\mp}\mathcal{K}_{\pm})\Delta_{\pm} \!\Bigr)+4(E')^2\nu_{2}\mathrm{tr}(v_{\pm}^2) \, .
\end{equation}
The last term on the right hand side of \eqref{vpm2} provides a first hint of recursive structures, while the middle term has the same form as the first term on the right hand side of \eqref{d} and should be analysed separately. Using again \eqref{b} and \eqref{a} one easily finds that  
\begin{equation}\label{e}
\begin{aligned}
\mathrm{tr}\Bigl(\!(\mathcal{O}_{\mp}\mathcal{K}_{\pm})\Delta_{\pm} \!\Bigr)
\deq&
-2E' \mathrm{tr}(v_{\pm}^2)\mathrm{tr} \Bigl(\!(\mathcal{O}_{\mp}\mathcal{K}_{\pm})(\mathcal{O}_{\mp}\mathcal{K}_{\mp})  \!\Bigr)
\\
&+4(E')^2\nu_{2}\mathrm{tr}\Bigl(\! (\mathcal{O}_{\pm}\mathcal{K}_{\pm})^2 \!\Bigr)
+4(E')^2\nu_{2}\mathrm{tr}\Bigl( (\mathcal{O}_{\mp}\mathcal{K}_{\pm})\Delta_{\pm}  \Bigr) \, ,
\end{aligned}
\end{equation}
which finally closes the circle of sought relations. At this point, given the interaction function $E(\nu_2)$ and its derivative $E'(\nu_2)$, one can first substitute \eqref{d} in \eqref{c} and after exploiting the definition $\nu_{2}:=\mathrm{tr}(v_{+}^2)\mathrm{tr}(v_{-}^2)$ recursively susbstitute the relations \eqref{vpm2} and \eqref{e} in the Lagrangian, expanding the resulting expression up to the desired order in $\lambda_{2k}$ and making sure that no $v_{\pm}$ and $\Delta_{\pm}$ appear in the final truncated expression. The procedure can be iterated to any order in $\lambda_{2k}$ and in principle provides the full deformed AFSM Lagrangian for any model in the class \eqref{AF-general-EOM-for-identity}, independently of the underlying Lie algebra structure. As an example, we report here the resulting expression up to $O(\lambda_{2k}^3)$
\begin{equation}\label{deformed-lagrangian-2k-flows}
\begin{aligned}
\mathcal{L}(\phi) \deq& -\frac{1}{2} \mathrm{tr} (\mathcal{K}_{+}\mathcal{O}_{-}\mathcal{K}_{-})
\\
& +\lambda_{2k}\mathrm{tr}\Bigl(\!(\mathcal{O}_{-}\mathcal{K}_{+})^2 \!\Bigr)^k\mathrm{tr}\Bigl(\!(\mathcal{O}_{-}\mathcal{K}_{-})^2\!\Bigr)^k
\\
& -4k^2\lambda_{2k}^2\mathrm{tr}\Bigl(\! (\mathcal{O}_{-}\mathcal{K}_{+})^2\!\Bigr)^{2k-1}\mathrm{tr}\Bigl(\!(\mathcal{O}_{-}\mathcal{K}_{-})^2 \!\Bigr)^{2k-1}\mathrm{tr}\Bigl(\! (\mathcal{O}_{-}\mathcal{K}_{+})(\mathcal{O}_{-}\mathcal{K}_{-})\!\Bigr)
\\
& -4k^3\lambda_{2k}^3 \mathrm{tr}\Bigl(\! (\mathcal{O}_{-}\mathcal{K}_{+})^2\!\Bigr)^{3k-2}\mathrm{tr}\Bigl(\!(\mathcal{O}_{-}\mathcal{K}_{-})^2\!\Bigr)^{3k-2} \cdot
\\
& \quad  \cdot \Bigl[(k-2) \mathrm{tr}\Bigl(\!(\mathcal{O}_{-}\mathcal{K}_{+})^2\!\Bigr)\mathrm{tr}\Bigl(\!(\mathcal{O}_{-}\mathcal{K}_{-})^2\!\Bigr)-4(2k-1)\mathrm{tr}\Bigl(\!(\mathcal{O}_{-}\mathcal{K}_{+})(\mathcal{O}_{-}\mathcal{K}_{-})\!\Bigr)^2 
\Bigr] \, .
\end{aligned}
\end{equation}
An important thing to notice is that while the even higher spin currents \eqref{ansatz-even-flows} and the associated interaction function \eqref{E-f-g-spin-2k} only depend on chiral traces of auxiliary fields via $\nu_{2}$, at order $\lambda_{2k}^2$ the deformed Lagrangian starts to exhibit contributions involving non-chiral traces of the fundamental 
fields, which can be written as
\begin{equation}\label{non-chiral-traces-2k-flows}
\mathrm{tr}\Bigl(\! (\mathcal{O}_{-}\mathcal{K}_{+})(\mathcal{O}_{-}\mathcal{K}_{-})\!\Bigr) = \mathrm{tr}(\mathcal{K}_{+}\mathcal{K}_{-})+\mathrm{tr}\Bigl(\mathcal{K}_{+} \, \frac{M^2}{1-M^2} \, \mathcal{K}_{-}\Bigr) \, ,
\end{equation}
after recalling the general form \eqref{operators-general-form} of $\mathcal{O}_{\pm}$, which implies that 
\begin{equation}\label{OpOm-2k-flows}
\mathcal{O}_{+}\mathcal{O}_{-}=\frac{1}{1-M^2} = 1+\frac{M^2}{1-M^2} \, .
\end{equation}
Generically, in a setting without auxiliary fields, it is quite non-trivial to take into account these contributions when constructing an ansatz for the higher-spin currents and/or the deformed Lagrangian, since the possible number of terms that the ansatz should include rapidly proliferates when no assumption on the chirality is made. The auxiliary field construction seems to automatically take into account such contribution and while this may initially sound like black magic, a closer look at the equations above actually reveals how all such terms are truly build into the formalism: the variations $\Delta_{\pm}$ of the interaction function, given in \eqref{b}, are indeed proportional to $v_{\mp}$ and this immediately implies that the EOM \eqref{a} relate $v_{\pm}$ at $O(\lambda_{2k}^0)$ to $v_{\mp}$ at higher orders. Consecutively, even starting with an ansatz that consists of purely chiral terms $\mathrm{tr}(v_{\pm}^{k})$, when recursively exploiting the EOM -- or equivalently the relations \eqref{vpm2} and \eqref{e} -- to integrate out the auxiliary fields, one automatically generates all possible combinations of non-chiral traces at different orders in the expansion parameter. This feature holds true in the construction of both the deformed Lagrangian and the conserved currents, when expressed in terms of fundamental fields, and represents another crucial advantage of the auxiliary field construction. While this was not apparent in the free boson case, due to the lack of Lie algebraic structures, it immediately became clear in the simplest case of AFSM involving only $\nu_{2}$, and will also reappear in the more complicated setting of section \ref{sec:su3}, when integrating out the auxiliary fields in \ref{sec:su3-deformed-lagrangian}.

Another interesting remark is that the deformed Lagrangian \eqref{deformed-lagrangian-2k-flows} reduces to the one obtained for the single free boson \eqref{O(4)-boson-lagrangian} -- with $n=2k$ -- after setting $\mathcal{O}_{\pm}=\mathds{1}$,  $\mathcal{K}_{\pm}=\partial_{\pm}\phi$ and discarding the traces via the naive substitution $\mathrm{tr}(X)\rightarrow X$. While curious, also this property is in fact already encoded in the whole construction, which relies on the formal substitution of $\nu:=v_+v_-$ with $ \sqrt{\nu_2}=:\sqrt{\mathrm{tr}(v_+^2)\mathrm{tr}(v_-^2)}$, that becomes an identity under the above naive removal of the traces.  

In conclusion, we see that we may accomplish our second goal and characterize all solutions of even-spin Smirnov-Zamolodchikov flow equations (\ref{flow}) in a fairly large class of models -- namely, any auxiliary field model obeying (\ref{AF-general-EOM-for-identity}) and with an interaction function $E ( \nu_2 )$ depending only on $\nu_2$ -- by essentially reducing the problem to the one that we have already solved for the simpler case of the single free boson.

\section{Spin-$3$ flows of AF sigma models based on $\mathfrak{su}(3)$}
\label{sec:su3}

In this section, we will again consider the second problem posed in our paper -- finding triplets $(E, \tau_{\pm n}, \theta_{\pm ( n - 2 )})$ which describe solutions to Smirnov-Zamolodchikov flows -- in one additional example, which is different from the scenario of Section \ref{sec:even_higher_spin} because the flow is associated with an \emph{odd} value of $n$. We will study SZ flows driven by spin-$3$ conserved currents in examples based on a specific choice of underlying Lie algebraic structure, namely $\mathfrak{su}(3)$. Once again, our construction will solely rely on the requirements \eqref{AF-general-EOM-for-identity} and the resulting identity \eqref{important-identity}, hence implying a general validity for any AF sigma model meeting the necessary requirements. As partially anticipated in the review section \ref{section:review}, this choice of underlying Lie algebraic structure will play the important role of making truly manifest the need for an enlarged ansatz for the interaction function, such that the old recipe \cite{Bielli:2024ach,Bielli:2024fnp,Bielli:2024oif}
\begin{equation}
E(v):=E(\nu_{2},...,\nu_{N}) 
\qquad
\text{with} 
\qquad \nu_{k}:=\mathrm{tr}(v_{+}^k)\mathrm{tr}(v_{-}^k) \qquad \forall \,  k=2,...,N \, ,
\end{equation}
exhibiting connection to the Smirnov-Zamolodchikov flows to leading order in the deformation parameter, needs to be extended to 
\begin{equation}
E(v):=E(\nu_{+2},...,\nu_{+N},\nu_{-2},...,\nu_{-N}) 
\quad \text{with} \quad
\nu_{\pm k}:=\mathrm{tr}(v_{\pm}^k) 
\quad \forall \, k\in \{2,...N\} \, ,
\label{newE_again}
\end{equation}
for this connection to be still valid beyond the leading order. We stress again that classical integrability is preserved by all interaction functions in this new family, since the condition
\begin{equation}
[\Delta_{\pm}, v_{\mp}] \deq 0
\qquad \text{with} \qquad 
\Delta_{\pm}:= \delta_{v_{\mp}}E(v) 
\end{equation}
remains unaffected -- see comments around equation \eqref{Delta-enlarged-ansatz} -- and the requirement of Lorentz invariance translates into the conditions \eqref{homogeneity-condition}.

The idea is to follow the same strategy used in previous sections: reducing the conservation equation for some higher-spin currents to a system of PDEs in Lorentz invariant variables, solving then perturbatively beyond leading order in the deformation parameter, together with the Smirnov-Zamolodchikov flow. For concreteness, and viability of the analysis, we will restrict ourselves to the case of spin-3 currents. Again, since in this section we do not concern ourselves with the construction of higher-spin currents for \emph{generic} interaction functions $E$ (the first goal mentioned in the introduction), focusing only on the solutions to spin-$3$ Smirnov-Zamolodchikov flow equations (the second goal), we will freely use the assumption that $E$ obeys a differential equation (\ref{flow}) to simplify the analysis.

\subsection{Need for new variables}
There are three key observations that we have been using in the previous sections in order to solve the spin-$n$ free boson flows first, and the spin-$2k$ AFSM flows later:

\begin{enumerate}
    \item\label{pointone} The connection to SZ flows established in previous works, reported in equation \eqref{E-desired-behaviour}, was exploited in the free boson case to restrict the leading order behaviour of the interaction function to $E\simeq \lambda_{n} \, \nu^n $, and then solve conservation and flow equations.
    \item\label{pointtwo} Using the unified framework in section \ref{sec:unification} we showed that the 4 derivatives $\del_\pm \,\ttr {v_+ ^2}$ and $\del_\pm \,\ttr {v_- ^2}$ are related by algebraic equations \textit {only} involving $\ttr {v_\pm^2}$. 
    \item Point \ref{pointtwo} implies that it is consistent to reduce the conservation equation for general AFSM deformed by $E=E(\nu _2)$ to a system of ODEs in $\nu _2$, which also closely resembles the free boson case and can be solved using the results of point \ref{pointone} after the mapping.
\end{enumerate}
In the spin-3 case, point \ref{pointone} leads to $E\simeq \lambda_{3} \, \nu _3$ and it would thus be tempting to expect that 
\begin{equation}
    E = E(\nu _2 ,\nu _3)\, ,
\end{equation}
would lead, via point \ref{pointtwo}, to a conservation that can be reduced to a system of PDEs in $\nu _2$ and $\nu _3$. However, as anticipated, we will see that this is not quite correct, since two more Lorentz invariant variables built in terms of $\ttr {v_\pm ^2}$ and $\ttr {v_\pm ^3}$ must be added among the dependences of the interaction function. To begin, the derivative identities relating 
\begin{equation}
    \del _\pm \ttr {v_+ ^2 } 
    \quad , \quad 
    \del _\pm \ttr {v_+ ^3}
    \quad , \quad 
    \del _\pm \ttr {v_- ^2 }
    \quad ,  \quad 
    \del _\pm \ttr {v_- ^3}
\end{equation}
can be derived using \eqref{important-identity} and are written down in appendix \ref{app:su3der}. The details of the calculation are irrelevant; the only important observation is that   in this case we have 8 derivatives of traces that we can write down,
 and 4 algebraic equations relating them to each other. Thus, we expect a system of $8-4=4$ coupled PDEs. This also suggests that in order not to have an over-determined system, we should define the currents in terms of 4 functions of Lorentz invariant variables. Let us denote the spin-3 and spin-1 currents by $\tau _{\pm3}$, $\theta _{\pm1}$ respectively. One can think about how to construct a general product of chiral traces with spin $+3$
 \begin{equation}\label{eq:X+++}
    X_{+++}= \ttr {v_+^3}^p\ttr {v_-^3}^q\ttr{v_+^2}^r\ttr {v_-^2}^s
    \,,
 \end{equation}
 where all the integer powers are non-negative. Suppose $p=1+k$, then the combination\\ $\ttr {v_+^3}^k\ttr {v_-^3}^q\ttr{v_+^2}^r\ttr {v_-^2}^s$ must have spin 0.
    \begin{enumerate}
        \item Suppose that $k\geq q$, then $s\geq r$, and \eqref{eq:X+++} reduces to 
        \begin{equation}
            X_{+++}= \ttr {v_+^3}\nu _3^a\, \nu _2^b\rb{\ttr{v_+^3}^2\ttr{v_-^2}^3}^c\,.
        \end{equation}
        \item Now assume $k<q$ then $s<r$ and we must have 
        \begin{equation}
            X_{+++}= \ttr {v_+^3}\nu _3^a\, \nu _2^b\rb{\ttr{v_-^3}^2\ttr{v_+^2}^3}^c\,.
        \end{equation}
    \end{enumerate}
 Crucially, we see that there are infinitely many possible combinations which are not related by functions of $\nu _{2}$ and $\nu_{3}$, namely 
 \begin{equation}
     X^{(n)}_{+++}= \ttr {v_+^3}\rb{\ttr{v_+^3}^2\ttr{v_-^2}^3}^n\,.
 \end{equation}
 This suggests that in order to have a complete ansatz for the currents, which only involves finitely many function, we should include two more Lorentz invariant variables:
 \begin{equation}
     \omega = \ttr {v_+^3}^2\ttr{v_-^2}^3 
     \qquad ,\qquad 
     \widetilde \omega = \ttr {v_-^3}^2\ttr{v_+^2}^3\,.
 \end{equation}
 Note that these two variables are not independent, because 
 \begin{equation}
     \omega \, \widetilde \omega = \nu _3^2\,\nu _2^3\,,
 \end{equation}
 however, it will still be convenient to treat them as formally independent in what follows. With this modification in mind, we write the currents as
 \begin{equation}\label{eq:su3currents}
    \begin{split}
        \tau _{+3}&= f(\nu_2,\nu_3,\omega,\widetilde\omega) \, \ttr {v_+^3}+ g(\nu_2,\nu _3,\omega,\widetilde\omega)\, \ttr {v_-^3}\, \ttr {v_+^2}^3\,,\\[0.5em]
        \theta _{+1}&= h(\nu _2,\nu _3,\omega,\widetilde\omega)\, \ttr {v_+^3}\,\ttr {v_-^2}+ k(\nu _2,\nu _3,\omega,\widetilde\omega)\ttr{v_-^3}\ttr{v_+^2}^2\,.
    \end{split}
 \end{equation}
    Again, the expression above includes some redundancy, for example 
 \begin{equation}\label{eq:redundancy}
    \nu _3 \, \ttr {v_-^3 }\, \ttr {v_+^2}^3=\ttr {v_+^3}\, \widetilde \omega\,, 
 \end{equation}
nonetheless it will still be useful to parametrise functions as in \eqref{eq:su3currents}. \\

We will be interested in solving a Smirnov-Zamolodchikov flow sourced by spin-$3$ currents, and we have just shown that the relevant currents must involve some extra Lorentz invariant variables. We are then forced to include dependence on $\omega$ and $\widetilde \omega $ in the function $E$ as well. Importantly, deformations which include these extra variables still preserve integrability due to the simple equation \cite{Bielli:2024ach,Bielli:2024oif,Ferko:2024ali}
\begin{equation}
    [\Delta_\pm,v_\mp]\deq0 \, ,
\end{equation}
where $\Delta_{\pm}$ now includes contributions due to $E_{\omega}$ and $E_{\widetilde\omega}$.

\subsection{Series solution for the spin-$3$ flow}\label{sec:solution-spin-3-flow}
Armed with \eqref{eq:su3derid} and \eqref{eq:su3currents}, we can now construct a series solution order by order in the coupling $\lambda_{3}$ for both the currents and the interaction function $E$. The coupled system is
\begin{equation}\label{eq:su3flow}
    \begin{split}
        \del_- \, \tau_{+3}+\del _+\, \theta _{+1}&=0\,,\\[0.4em]
        \tau _{+3}\, \tau _{-3}-\theta _{+1}\,\theta_{-1}&=\del _{\lambda_{3}} E\,.
    \end{split}
\end{equation}
Unlike in the free boson case the system of PDEs arising from \eqref{eq:su3flow} is extremely complicated. Since what we are seeking is a series solution, we instead write down arbitrary expansions in the coupling $\lambda_{3}$ and solve order by order. The series expansion is greatly simplified by some dimensional analysis. We know from \cite{Bielli:2024ach} that to leading order in $\lambda_{3}$
\begin{equation}
    E= \lambda_{3} \, \nu _3 + \ldots 
\end{equation}
so, as discussed around equation \eqref{dim-analysis3}, one has $[\lambda_{3} ]=-4$ and
\begin{equation}
[\nu_{2}]=4 
\qquad , \qquad
[\nu_{3}]=6
\qquad , \qquad 
[\omega]=[\tilde{\omega}]=12 \, .
\end{equation}
The flow equation in \eqref{eq:su3flow} implies the following dimensions for the Lorentz invariant functions that appear in \eqref{eq:su3currents}
\begin{equation}
[f]=0 
\qquad , \qquad 
[g]=-6
\qquad , \qquad 
[h]=-2
\qquad , \qquad 
[k]=-4 \, .
\end{equation}
Accordingly, the functions which define \eqref{eq:su3currents} take the form 
\begin{equation}\label{eq:su3curr}
    \begin{split}
        f &= f(\lambda_{3}\nu_2,\lambda_{3}^3\nu_3^2,\lambda_{3}^3\omega,\lambda_{3}^3\widetilde{\omega})\,, \\[0.4em]
        g&=\lambda_{3} ^3 \nu _3 \, G(\lambda_{3}\nu_2,\lambda_{3}^3\nu_3^2,\lambda_{3}^3\omega,\lambda_{3}^3\widetilde{\omega})\,, \\[0.4em]
        h&= \lambda_{3} ^2 \nu _3 H(\lambda_{3}\nu_2,\lambda_{3}^3\nu_3^2,\lambda_{3}^3\omega,\lambda_{3}^3\widetilde{\omega})\,,\\[0.4em]
        k&= \lambda_{3} \, K(\lambda_{3}\nu_2,\lambda_{3}^3\nu_3^2,\lambda_{3}^3\omega,\lambda_{3}^3\widetilde{\omega})\,.
    \end{split}
 \end{equation}
Similarly, the function $E$ is also constrained to 
\begin{equation}\label{su3-interaction-function-ansatz}
      E = \lambda_{3}\nu_3 {\mathcal{E}}(\lambda_{3}\nu_2,\lambda_{3}^3\nu_3^2,\lambda_{3}^3\omega,\lambda_{3}^3\widetilde{\omega})\,,
\end{equation}
on dimensional grounds. We further assume that the dimensionless functions $f,\, G,\, H,\,K,\, {\cal E}$ are analytic. Interestingly, we find that restricting to this class of functions makes $g$ completely spurious in light of equation \eqref{eq:redundancy}. Notably, we now have an overdetermined system of 4 equations with 3 unknowns. Nonetheless, we find by expanding in $\lambda_{3}$ that two out of the 4 equations are equivalent. In order to obtain the currents of opposite charge, namely $\tau _{-3}$ and $\theta _{-1}$, we flip all signs in $\tau_{+3}$ and $\theta _{+1}$ and swap $\omega \leftrightarrow \widetilde \omega $ in \eqref{eq:su3curr}. Finally, the series expansion up to $O(\lambda_{3} ^7)$ gives 
\begin{equation}\label{eq:SU3exp}
    \begin{split}
        \tau _{+3}&=\mathrm{tr}(v_+^3)\bigg(1-\lambda_{3} ^2 \,\frac9{16}\,\nu _2 ^2+ \lambda_{3} ^4\,\frac {81}{256}\,\nu_2\rb{16\,\widetilde \omega+\nu _2 ^3+ 8 \,\omega } 
        \\
        &
        +\lambda_{3} ^6\, \frac {81}{4096}\,\rb{576\,\widetilde \omega ^2+ 96\,\nu _2^3[7\,\widetilde\omega +48\,\nu _3^2]+ 9 \,\nu _2 ^6+296\, \nu _2 ^3\,\omega + 192 \, \omega ^2}\bigg)+ O(\lambda_{3} ^8)\,,
        \\
        \\
        \theta _{+1}
        &
        =\mathrm{tr}(v_-^3)\,\mathrm{tr}(v_+^2)^2\bigg (\!\!-\frac 34 \,\lambda_{3} + \frac {27}{64}\,\lambda_{3} ^3\, \nu _2^2 + \frac{81}{4} \,\lambda_{3} ^5\bigg [\frac3{32}\, \nu _2 \widetilde \omega +  3\,\nu _2 \, \nu _3^2 + \frac3{32}\, \nu _2 ^4 +\frac78\, \nu_2 \, \omega\bigg] 
        \\[0.4em]
        &
        +  \frac{243}{8} \lambda_{3}^7 \bigg [ \frac{3}{32} \widetilde \omega ^2 \!+\! \frac 9{2} \nu _3 ^2 \widetilde \omega \!+\! \frac{9}{2048} \nu _2 ^6 \!+\! \frac {57}{64} \nu _2 ^3 \omega \!+\! \frac {81}{16} \omega ^2 \!+\! \frac {37}{256} \nu _2 ^3 \widetilde \omega \!+\! \frac {207} {16}  \nu _2 ^3 \nu _3 ^2 \!+\! 27 \nu _3 ^2  \omega  \bigg ] \bigg)
        \\[0.4em]
        &
        + \mathrm{tr}(v_+^3 )\mathrm{tr}(v_-^2)\bigg (\frac{27}4 \,\lambda_{3} ^3\, \nu _3 \, \nu _2 +\frac {243}{16}\,\lambda_{3} ^5\, \nu _3 \, \omega \bigg)+ O(\lambda_{3} ^8)\,,
        \\
        \\
        E
        & = \lambda_{3} \, \nu _3 \bigg (1- \frac 9{16} \, \lambda_{3} ^2 \, \nu _2 ^3  + \frac {81}{256}\, \lambda_{3} ^4  \nu _2 \big (8 \, [\widetilde \omega +\omega ]+ \nu _2 ^3  \big)\\[0.4em]
        &
        + \frac {81}{4096}\, \lambda_{3} ^6 \big (192 [\widetilde \omega ^2+ \omega^2 ]+ 296\, \nu _2 ^3 \,[ \widetilde \omega +\omega]+2304\, \nu _2 ^3 \, \nu _3 ^2 + 9 \, \nu _2 ^6  \big )\bigg )
        + O(\lambda_{3} ^8)\,.
    \end{split}
\end{equation}
It is worth pointing out that \eqref{eq:SU3exp} is not the unique way to write the currents and $E$. Again this is because $\omega$ and $\widetilde \omega $ are not independent. However, this ambiguity disappears when everything is expressed explicitly in terms of traces. Note that the currents with opposite charge, namely $\tau _{-3}$ and $\theta_{-1} $, can be obtained by flipping all signs in \eqref{eq:SU3exp}.

\subsection{Deformed AF Lagrangian at $O(\lambda_{3}^2)$}\label{sec:su3-deformed-lagrangian}
We conclude our analysis of this instructive $\mathfrak{su}(3)$ example by integrating out, up to order $\lambda_{3}^2$, the auxiliary fields from any AF sigma models which can be framed in the language of section \ref{sec:spin-2k-of-AFSM}. This final result confirms the strength of the new unified viewpoint on AFSM and allows to appreciate once more the appearance of trace contributions which involve non-chiral products of the fundamental fields, even when starting from purely chiral traces of auxiliary fields. This feature was already observed in section \ref{sec:even-SZ-flows} for the simplest case of interaction functions depending on $\nu_{2}$ only, and represents a novel interesting strength of the auxiliary field formalism.
The starting point is the AFSM Lagrangian written in the form \eqref{common-Lagrangian}, which using the auxiliary fields EOM \eqref{common-v-EOM} can be written as \eqref{integrated-Lagrangian-step-0}. For simplicity we report here the two equations, specialising to the interaction function \eqref{su3-interaction-function-ansatz} under consideration 
\begin{equation}\label{v-EOM+Lagrangian-final-section}
\begin{aligned}
v_{\pm}+&\mathcal{O}_{\pm}(\mathcal{K}_{\pm})+\mathcal{O}_{\mp}^{^{-1}}\mathcal{O}_{\pm}\Delta_{\pm}\deq 0 \, ,
\\
\mathcal{L}(\phi,v) \deq  -\frac{1}{2} \mathrm{tr} (\mathcal{K}_{+}\mathcal{O}_{-}\mathcal{K}_{-})+&\mathrm{tr}(\Delta_{+}\mathcal{O}^{-1}_{+}\mathcal{O}_{-}\Delta_{-})+\lambda_{3}\nu_3 {\mathcal{E}}(\lambda_{3}\nu_2,\lambda_{3}^3\nu_3^2,\lambda_{3}^3\omega,\lambda_{3}^3\widetilde{\omega}) \, .
\end{aligned}
\end{equation}
Aiming at $O(\lambda_{3}^2)$ contributions, the truncation of the interaction function \eqref{eq:SU3exp} to be considered is very simple and one can easily extract $\Delta_{\pm}$:
\begin{equation}\label{E-and-Delta-final-section}
E \simeq \lambda_{3} \nu_{3} + O(\lambda_{3}^3) \, ,
\qquad \qquad
\Delta_{\pm} = 3\lambda_{3} \, \mathrm{tr}(v_{\pm}^3)v_{\mp}^{A}v_{\mp}^{B}d_{AB}{}^{C}T_{C} \, .
\end{equation}
The Lagrangian \eqref{v-EOM+Lagrangian-final-section} then becomes
\begin{align}\label{final-section-lagrangian-intermediate}
\mathcal{L} \!\deq&\! -\!\frac{1}{2}\mathrm{tr}(\mathcal{K}_{+}\mathcal{O}_{-}\mathcal{K}_{-})+\lambda_{3} \mathrm{tr}(v_{+}^3)\mathrm{tr}(v_{-}^{3})+
\\
& \! +\!9\lambda_{3}^2 \mathrm{tr}(v_{+}^3)\mathrm{tr}(v_{-}^{3})v_{+}^{A}v_{+}^{B}v_{-}^{D}v_{-}^{E}d_{AB}{}^{C}d_{DE}{}^{F}\mathrm{tr}(T_{F}\mathcal{O}_{+}^{-1}\mathcal{O}_{-}T_{C})
\notag\\
=&\!-\!\frac{1}{2}\mathrm{tr}(\mathcal{K}_{+}\mathcal{O}_{-}\mathcal{K}_{-})+\lambda_{3} \mathrm{tr}(v_{+}^3)\mathrm{tr}(v_{-}^{3})+
\notag\\
& \!+\!9\lambda_{3}^2 \mathrm{tr}\Bigl(\! (\mathcal{O}_{+}\mathcal{K}_{+})^3 \!\Bigr) \mathrm{tr}\Bigl(\! (\mathcal{O}_{-}\mathcal{K}_{-})^3 \!\Bigr)(\mathcal{O}_{+}\mathcal{K}_{+})^{A}(\mathcal{O}_{+}\mathcal{K}_{+})^{B}d_{AB}{}^{C}\mathrm{tr}\Bigl(\!(\mathcal{O}_{-}\mathcal{K}_{-})^2 \mathcal{O}_{+}^{-1}\mathcal{O}_{-}T_{C}\!\Bigr) \, ,
\notag
\end{align}
after having rearranged the $O(\lambda_{3}^2)$ trace term as
\begin{equation}
d_{AB}{}^{C}d_{DE}{}^{F}\mathrm{tr}(T_{F}\mathcal{O}_{+}^{-1}\mathcal{O}_{-}T_{C})= d_{AB}{}^{C}\mathrm{tr}(T_{(D}T_{E} \mathcal{O}_{+}^{-1}\mathcal{O}_{-}T_{C)})
\end{equation}
and substituted the auxiliary fields EOM \eqref{v-EOM+Lagrangian-final-section} at $O(\lambda_{3}^0)$. The final step is the extraction of $O(\lambda_{3}^2)$ terms from the double trace term on the first line of the Lagrangian, which still contains the auxiliary fields. Using the EOM \eqref{v-EOM+Lagrangian-final-section} and \eqref{E-and-Delta-final-section} recursively one finds
\begin{equation}
v_{\pm}\deq -\mathcal{O}_{\pm}\mathcal{K}_{\pm}+3\lambda_{3}\mathrm{tr}\Bigl(\! (\mathcal{O}_{\pm}\mathcal{K}_{\pm})^3 \!\Bigr)(\mathcal{O}_{\mp}\mathcal{K}_{\mp})^{A}(\mathcal{O}_{\mp}\mathcal{K}_{\mp})^{B}d_{AB}{}^{C}\mathcal{O}_{\mp}^{-1}\mathcal{O}_{\pm}T_{C} + O(\lambda_{3}^2) \, ,
\end{equation}
which immediately leads to
\begin{align}\label{final-section-double-trace-term}
& \mathrm{tr}(v_{+}^{3})\mathrm{tr}(v_{-}^{3})
\deq \mathrm{tr}\Bigl(\! (\mathcal{O}_{+}\mathcal{K}_{+})^3 \!\Bigr)\mathrm{tr}\Bigl(\! (\mathcal{O}_{-}\mathcal{K}_{-})^3 \!\Bigr) +
\\
& - 9\lambda_{3} \mathrm{tr}\Bigl(\! (\mathcal{O}_{+}\mathcal{K}_{+})^3 \!\Bigr)\mathrm{tr}\Bigl(\! (\mathcal{O}_{-}\mathcal{K}_{-})^3 \!\Bigr) (\mathcal{O}_{+}\mathcal{K}_{+})^{A}(\mathcal{O}_{+}\mathcal{K}_{+})^{B}d_{AB}{}^{C}\mathrm{tr}\Bigl(\! (\mathcal{O}_{-}\mathcal{K}_{-})^2\mathcal{O}_{+}^{-1}\mathcal{O}_{-}T_{C} \!\Bigr)+
\notag\\
&  - 9\lambda_{3} \mathrm{tr}\Bigl(\! (\mathcal{O}_{+}\mathcal{K}_{+})^3 \!\Bigr)\mathrm{tr}\Bigl(\! (\mathcal{O}_{-}\mathcal{K}_{-})^3 \!\Bigr) (\mathcal{O}_{-}\mathcal{K}_{-})^{A}(\mathcal{O}_{-}\mathcal{K}_{-})^{B}d_{AB}{}^{C}\mathrm{tr}\Bigl(\! (\mathcal{O}_{+}\mathcal{K}_{+})^2\mathcal{O}_{-}^{-1}\mathcal{O}_{+}T_{C} \!\Bigr) \, .
\notag
\end{align}
Substituting then \eqref{final-section-double-trace-term} into the $O(\lambda_{3})$ term in \eqref{final-section-lagrangian-intermediate} two terms cancel exactly and one is finally left with the desired $O(\lambda_{3}^2)$ Lagrangian
\begin{align}\label{su3-lagrangian}
\mathcal{L}\deq& -\frac{1}{2}\mathrm{tr}(\mathcal{K}_{+}\mathcal{O}_{-}\mathcal{K}_{-})+\lambda_{3} \mathrm{tr}\Bigl(\! (\mathcal{O}_{+}\mathcal{K}_{+})^3 \!\Bigr)\mathrm{tr}\Bigl(\! (\mathcal{O}_{-}\mathcal{K}_{-})^3 \!\Bigr) +
\\
& - 9\lambda_{3}^2 \mathrm{tr}\Bigl(\! (\mathcal{O}_{+}\mathcal{K}_{+})^3 \!\Bigr)\mathrm{tr}\Bigl(\! (\mathcal{O}_{-}\mathcal{K}_{-})^3 \!\Bigr) (\mathcal{O}_{-}\mathcal{K}_{-})^{A}(\mathcal{O}_{-}\mathcal{K}_{-})^{B}d_{AB}{}^{C}\mathrm{tr}\Bigl(\! (\mathcal{O}_{+}\mathcal{K}_{+})^2\mathcal{O}_{-}^{-1}\mathcal{O}_{+}T_{C} \!\Bigr) \, .
\notag
\end{align}
The appearance of non-chiral traces can then be observed by looking at the $O(\lambda_{3}^2)$ term in the Lagrangian -- as in section \ref{sec:even-SZ-flows}. Recalling the general form \eqref{operators-general-form} of $\mathcal{O}_{\pm}$ one finds
\begin{equation}
\mathcal{O}_{-}^{-1}\mathcal{O}_{+}=\frac{1-M}{1+M} = 1-\frac{2M}{1+M} \, ,
\end{equation}
such that the lowest order contribution in $M$ can be written as
\begin{equation}\label{non-chiral-traces-su3}
\begin{aligned}
(\mathcal{O}_{-}\mathcal{K}_{-})^{A}&(\mathcal{O}_{-}\mathcal{K}_{-})^{B}d_{AB}{}^{C}\mathrm{tr}\Bigl(\! (\mathcal{O}_{+}\mathcal{K}_{+})^2T_{C} \!\Bigr) =
\\
&= (\mathcal{O}_{-}\mathcal{K}_{-})^{A}(\mathcal{O}_{-}\mathcal{K}_{-})^{B}(\mathcal{O}_{+}\mathcal{K}_{+})^{D}(\mathcal{O}_{+}\mathcal{K}_{+})^{E} d_{AB}{}^{C}d_{DEC}
\\
& = \mathrm{tr} \Bigl(\! (\mathcal{O}_{+}\mathcal{K}_{+})^2(\mathcal{O}_{-}\mathcal{K}_{-})^2 \!\Bigr)-\frac{1}{36}\mathrm{tr} \Bigl(\! (\mathcal{O}_{+}\mathcal{K}_{+})^2 \!\Bigr)\mathrm{tr} \Bigl(\! (\mathcal{O}_{-}\mathcal{K}_{-})^2 \!\Bigr) \, ,
\end{aligned}
\end{equation}
after having used the identity
\begin{equation}
J_{+}^{A}J_{+}^{B}J_{-}^{D}J_{-}^{E}d_{AB}{}^{C}d_{DEC} = \mathrm{tr}(J_{+}^2J_{-}^2)-\frac{1}{36}\mathrm{tr}(J_{+}^2)\mathrm{tr}(J_{-}^2) \qquad \text{for any} \qquad J\in \mathfrak{su}(3) \, ,
\end{equation}
then the $O(\lambda_{3}^2)$ Lagrangian becomes
\begin{align}
\mathcal{L} \!\deq&\! -\!\frac{1}{2}\mathrm{tr}(\mathcal{K}_{+}\mathcal{O}_{-}\mathcal{K}_{-}) \!+\! \lambda_{3} \mathrm{tr}\Bigl(\! (\mathcal{O}_{+}\mathcal{K}_{+})^3 \!\Bigr)\mathrm{tr}\Bigl(\! (\mathcal{O}_{-}\mathcal{K}_{-})^3 \!\Bigr) 
\\
& \!-\! 9\lambda_{3}^2 \mathrm{tr}\Bigl(\! (\mathcal{O}_{+}\mathcal{K}_{+})^3 \!\Bigr)\mathrm{tr}\Bigl(\! (\mathcal{O}_{-}\mathcal{K}_{-})^3 \!\Bigr)\Bigl[\mathrm{tr} \Bigl(\! (\mathcal{O}_{+}\mathcal{K}_{+})^2(\mathcal{O}_{-}\mathcal{K}_{-})^2 \!\Bigr) \!-\! \frac{1}{36}\mathrm{tr} \Bigl(\! (\mathcal{O}_{+}\mathcal{K}_{+})^2 \!\Bigr)\mathrm{tr} \Bigl(\! (\mathcal{O}_{-}\mathcal{K}_{-})^2 \!\Bigr) \Bigr] 
\notag \\
& \!+\! 18 \lambda_{3}^2 \mathrm{tr}\Bigl(\! (\mathcal{O}_{+}\mathcal{K}_{+})^3 \!\Bigr)\mathrm{tr}\Bigl(\! (\mathcal{O}_{-}\mathcal{K}_{-})^3 \!\Bigr) (\mathcal{O}_{-}\mathcal{K}_{-})^{A}(\mathcal{O}_{-}\mathcal{K}_{-})^{B}d_{AB}{}^{C}\mathrm{tr}\Bigl(\! (\mathcal{O}_{+}\mathcal{K}_{+})^2\frac{M}{1\!+\!M}T_{C} \!\Bigr) \!\!+\! O(\lambda_{3}^3) \,.
\notag
\end{align}
This can then be easily specialised to each of the AFSM discussed in section \ref{sec:spin-2k-of-AFSM} by appropriately choosing the fundamental degrees of freedom $\mathcal{K}_{\pm}$ and operator $M$.\\

We conclude by noting that the $O(\lambda_{3}^2)$ non-chiral traces in \eqref{non-chiral-traces-su3} are more involved than the ones previously encountered in \eqref{non-chiral-traces-2k-flows} at the same order: this is due to the more complicated nature of the interaction function considered in this section, which does not only depend on $\nu_2$. What however remains almost unchanged, with respect to section \ref{sec:even-SZ-flows}, is the observation on how these new contributions arise in the deformed Lagrangian, as well as in the currents, when integrating out the auxiliary fields. In section \ref{sec:even-SZ-flows}, this was due to the fact that $\delta_{v_{\mp}}E(\nu_2)\propto v_{\mp}$, which combined with the EOM \eqref{common-v-EOM} caused the mixing at orders greater or equal than $2$ in the deformation parameter. Here the mechanism is the same, but the realisation is made more complex by the dependence on traces with higher powers of auxiliary fields. One can in fact make this observation a bit more precise by recalling the variations \eqref{Delta-enlarged-ansatz} for  an arbitrary interaction function in the new enlarged family \eqref{newE}:
\begin{equation}
\Delta_{\pm}:=\delta_{v_{\mp}}E(\nu_{+2},...,\nu_{+N},\nu_{-2},...,\nu_{-N})=\sum_{k=2}^{N} k \frac{\partial E}{\partial \nu_{\mp k}} v_{\mp}^{A_{1}}...v_{\mp}^{A_{k-1}}d_{A_{1}...A_{k-1}}{}^{B}T_{B} \, .
\end{equation}

Clearly $\Delta_{\pm}$ is still proportional to $v_{\mp}$ -- just in a much more complicated way involving multiple copies of it -- and for this reason the EOM \eqref{v-EOM+Lagrangian-final-section} still establishes a relation between $v_{\pm}$ at $O(\lambda_{3}^0)$ and $v_{\mp}$ at higher orders, which in turn generates non-chiral traces when recursively substituted into the desired expression to get rid of the auxiliary fields. This shows that the existence of non-chiral structures observed in this work is far from being an accident; it is rather an in-built feature of the auxiliary field formalism, which allows to avoid the encoding of such contributions in complicated general ans{\"a}tze.

\section{Conclusion}\label{sec:conclusion}

This work has addressed two open questions in the study of recently-introduced auxiliary field deformations of integrable $2d$ sigma models:
\begin{enumerate}[label=(\Roman*)]
    \item\label{goal_one_conc} What are the local, higher-spin conserved currents in these models?

    \item\label{goal_two_conc} Which models in this class satisfy higher-spin Smirnov-Zamolodchikov flows (\ref{flow})?
\end{enumerate}
In the process of answering these questions, we have also established two ``bonus'' results. The first is that the various families of auxiliary field models can be extended by allowing the interaction function $E$ to depend on a larger set of variables than the one described in \cite{Bielli:2024ach}, including ``mixed'' invariants like $\tr ( v_+^3 )^2 \tr ( v_-^2 )^3$ in addition to the usual variables $\nu_k = \tr ( v_+^k ) \tr ( v_-^k )$, while preserving integrability.\footnote{Not only are these new invariants \emph{allowed}, but in fact their inclusion is \emph{mandatory} if one would like to solve certain Smirnov-Zamolodchikov flows, such as the spin-$3$ flow of Section \ref{sec:su3}.} The second is that many of the auxiliary field sigma models that have been constructed thus far, including most of the ones in \cite{Ferko:2024ali,Bielli:2024ach,Bielli:2024fnp,Bielli:2024oif}, can all be written in the unified form described in Section \ref{sec:unification}, which makes it more transparent that these theories share a common structure and mechanism for integrability.

Regarding the primary goals \ref{goal_one_conc} and \ref{goal_two_conc}, in the simplest setting of AF deformations of a single free boson, we have proven that there always exists a solution to a set of differential equations which characterizes the local higher-spin conserved currents, and we have shown how to recursively determine the coefficients that define theories obtained from higher-spin SZ deformations. By exploiting the ``unified'' structure (\ref{AF-general-EOM-for-identity}) common to many auxiliary field models, we have also shown how to construct even higher-spin currents in a large class of auxiliary field models whose interaction functions depend only on $\nu_2$, and generate the solutions to Smirnov-Zamolodchikov flows in these cases. Finally, we have perturbatively studied SZ flows driven by spin-$3$ currents in a large family (again, any theories with the general structure (\ref{AF-general-EOM-for-identity})) of auxiliary field models with underlying $\mathfrak{su}(3)$ algebraic structure.

There remain several interesting directions for future research. One important set of questions concerns the definition of auxiliary field sigma models at the quantum level. At least for certain choices of interaction function, such as the one corresponding to the $\TT$ deformation or higher-spin Smirnov-Zamolodchikov flows beginning from the PCM, it is believed that the deformed theory should be well-defined quantum mechanically. For such cases that give rise to good quantum models, it is natural to ask whether the classically conserved higher-spin currents also persist in the quantum theory. This is by no means guaranteed; even for the simpler case of undeformed symmetric coset models, the question of whether classically conserved currents $T_{\pm \pm}^{n}$ built from components of the stress tensor remain conserved in the quantum theory has been studied in \cite{GOLDSCHMIDT1980392,Komatsu:2019hgc}, and is non-trivial. It would be interesting and useful to perform a similar investigation of the quantum properties of would-be higher-spin conserved currents in auxiliary field sigma models.

A second direction is to uncover the general structure underlying the integrability of the entire class of auxiliary field sigma models. Although the existing literature has constructed auxiliary field deformations on a case-by-case basis, starting from the PCM and then extending to its non-Abelian T-dual, Yang-Baxter and bi-Yang-Baxter deformations, etc., the structure (\ref{AF-general-EOM-for-identity}) seems to unify many of these cases and treat them within a single framework. It would be interesting to see whether one can crisply articulate the minimal set of assumptions about an integrable seed theory which are needed in order to ensure the existence of a family of auxiliary field deformations. Progress in this direction might allow us to apply such AF deformations to new theories, such as those which do not enjoy classical conformal invariance, or even to those which are not sigma models. For instance, finding a version of the AF formulation for integrable spin chains would be quite exciting.

Finally, it would be intriguing to investigate whether the auxiliary field formulation aids in finding a geometrical interpretation of higher-spin Smirnov-Zamolodchikov flows. As we mentioned in the introduction, for the case of spin-$2$ ($\TT$) deformations, the flows can be re-interpreted at the level of geometry: for instance, solutions to the deformed equations of motion can be mapped to solutions of the undeformed equations of motion on a field-dependent metric \cite{Conti:2018tca}. One might ask whether such an interpretation exists for more general auxiliary field deformations, both for interaction functions depending only on $\nu_2$ (i.e. general stress tensor deformations, which might all correspond to field-dependent metrics) and for those depending on other invariants, which might be understood via certain higher-spin generalizations, like a field-dependent $W$-metric (see e.g. \cite{Hull:1993kf} for an introduction). Such a result may also suggest a possible holographic interpretation of auxiliary field deformations. At least for the case of $\TT$, the field-dependent metric of \cite{Conti:2018tca} is closely related to the modified boundary conditions which holographically implement a boundary $\TT$ deformation in pure $\mathrm{AdS}_3$ gravity \cite{Guica:2019nzm}; these modified boundary conditions can also be described in the language of $SL(2 , \mathbb{R} ) \times SL ( 2, \mathbb{R} )$ Chern-Simons \cite{Llabres:2019jtx,Ouyang:2020rpq,Ebert:2022ehb}. It would be very interesting if higher-spin SZ flows can likewise be understood holographically via modified boundary conditions for Chern-Simons with gauge group $SL(N , \mathbb{R} ) \times SL ( N, \mathbb{R} )$.

\section*{Acknowledgements}

We are very grateful to Nicola Baglioni, Nicolò Brizio, Jue Hou, Zejun Huang, Tommaso Morone, Parita Shah, Liam Smith, and Roberto Tateo for discussions and collaborations related to this project as well as to all the participants of the ``Deformations of Quantum Field and Gravity Theories'' mini-workshop, held at the University of Queensland (UQ) from January 30 - February 6, 2025, for many productive discussions about the ideas in this paper and related topics. C.\,F. and D.\,B. are also happy to thank the University of Queensland for partial financial support to attend this mini-workshop, where part of this work was carried out.
D.\,B. is supported by Thailand NSRF via PMU-B, grant number B13F680083.
C.\,F. is supported by the National Science Foundation under Cooperative Agreement PHY-2019786 (the NSF AI Institute for Artificial Intelligence and Fundamental Interactions).
M.\,G. and G.\,T.-M. have been supported by the Australian Research Council (ARC) Future Fellowship FT180100353, ARC Discovery
Project DP240101409, the Capacity Building Package at the University of Queensland and a faculty start-up funding of UQ's School of Mathematics and Physics.

\newpage

\appendix

\section{ODEs for the AF free boson}\label{appendix:ODEs}
Here we describe some interesting features of the system \eqref{system-original-form}, reported below for clarity
\begin{equation}\label{system-original-form-appendix}
\begin{aligned}
0&=f' + nf + \frac{\nu g' + (n-2)g}{\nu^2 E''} - \frac{f'(\nu E')'}{E''}  \, , 
\\
0&=g'- (\nu E')' \frac{\nu g' + (n-2)g}{\nu^2 E''} + \frac{f'}{E''} \, .
\end{aligned}
\end{equation}
Despite a seemingly innocuous aspect, which might possibly be associated with the simplicity of the physical model from which it is derived -- namely the AF free boson -- the system \eqref{system-original-form-appendix} is in fact extremely complicated to solve in full generality, i.e. for $f$ and $g$ without specifying $E$. This has been shown in Theorem \ref{nogo_thm}, which states the impossibility of solving \eqref{system-original-form-appendix} with any ansatz for $f$ and $g$ involving finitely many derivatives of an arbitrary interaction function $E$. While the lack of a closed-form solution was overcome in section \ref{subsec:perturbative-analysis-free-boson} by requiring $f,g,E$ to be analytic, here we will report some interesting manipulations which can be applied to \eqref{system-original-form-appendix}, revealing some quite intriguing structures.

\subsection*{Some possible solutions and rearrangements}
We begin by rewriting the system   \eqref{system-original-form-appendix} as
\begin{equation}\label{system-first-form}
\begin{aligned}
f' & = \frac{n E' E''}{[(E')^2-1]} f - \frac{(n-2)E''}{\nu[(E')^2-1]} g \, , 
\\
g' &= \frac{n\nu E''}{[(E')^2-1]} f - \frac{(n-2)[(E')^2-1+\nu E'E'']}{\nu [(E')^2-1]} g \, ,
\end{aligned}
\end{equation}
and noting that
\begin{itemize}
\item As discussed above theorem \ref{nogo_thm}, the case of spin-2 currents corresponds to the stress-energy tensor. The system knows this information and indeed setting $n=2$ dramatically simplifies the equations, which can be integrated leading to \eqref{stress_tensor_boson}
\begin{equation}\label{system-for-spin-2}
\begin{cases}
   f' & = \frac{2 E' E''}{[(E')^2-1]} f 
\\
g' &= \frac{2\nu E''}{[(E')^2-1]} f  
\end{cases}
\qquad 
\Rightarrow 
\qquad 
\begin{cases}
f(\nu) &=c_{1}[(E'(\nu))^2-1]
\\
g(\nu) &= c_{2} + 2c_{1}[\nu E'(\nu)-E(\nu)] 
\end{cases} \, ,
\end{equation}
with $c_{1},c_{2}$ integration constants. The dimensional analysis around \eqref{dim-analysis3} and the ansatz \eqref{free-bosono-higher-spin-currents-guess} imply $[f]=0$ and $[g]=2$, so that one has
\begin{equation}\label{dimension-summary1}
[E]=[\nu]=[g]=2
\,\,\,\,\,\,\,\, , \,\,\,\,\,\,\,\, 
[f] =[E']=[g']= 0 
\,\,\,\,\,\,\,\, , \,\,\,\,\,\,\,\, 
[E''] = [f'] = -2 \, .
\end{equation}
In turn, $[c_{1}]=0$ and $[c_{2}]=2$, such that $c_{1}$ is a number and $c_{2}\propto \tfrac{1}{\lambda_{2}}$, since $[\lambda_{2}]=-2$. 
\item While for $n\neq 2$ there is no closed-form solution, one can try to solve \eqref{system-first-form} requiring $E$ to be of a prescribed form. For example, recalling $[\lambda_k]=2-2k$, one can find
\begin{align}
&
\!\!\!\!\!\!\!\!
\begin{cases}
E \!=\! \lambda_{2}\nu^{2} 
\qquad , \qquad f \!=\! c_{1} H_{2F1}[n-1,-\tfrac{n}{2},\tfrac{n}{2},4\lambda_{2}^2 \nu^2]
\,,
\\
g \!=\! 2\lambda_{2}\nu^2\frac{n}{(n-2)}f+4c_{1}\lambda_{2}\nu^2(4\lambda_2^2\nu^2-1)\frac{(n-1)}{(n-2)}H_{2F1}[1-\tfrac{n}{2},n,1+\tfrac{n}{2},4\lambda_{2}^2 \nu^2]
\,,
\end{cases}
\\
\notag \\
&
\!\!\!\!\!\!\!\!
\begin{cases}
E \!=\! \lambda_{3}\nu^{3}
\qquad , \qquad 
f \!=\! c_{1} H_{2F1}[\frac{3(n-1)}{4},-\tfrac{n}{2},\tfrac{n+1}{4},9\lambda_{3}^2 \nu^4]
\,,
\\
g \!=\! 3\lambda_{3}\nu^3\frac{n}{(n-2)} f +9c_{1}\lambda_{3}\nu^3(9\lambda_{3}^2\nu^4-1)\frac{n(n-1)}{(n-2)(n+1)} H_{2F1}[1-\tfrac{n}{2},\!\tfrac{1+3n}{4}\!,\!\tfrac{5+n}{4}\!,9\lambda_{3}^2\nu^4]
\,.
\end{cases}
\end{align}
Similar solutions of the form $E=\lambda_{k}\nu^{k}$ for generic $k$ should reasonably exist, but we were not able to obtain them using Mathematica. In principle, collecting a few more data points, such as solutions for $k=4,5,6$ one might be able to spot some pattern in the arguments of the above Gaussian hypergeometric functions ($H_{2F1}$), so as to make a guess for $f,g$ at generic $k$ and check whether or not it satisfies the equations. 
Finally, as one would expect, there exist solutions for non-analytic interaction functions like
\begin{equation}\label{non-analytic-solutions}
E=\frac{1}{\lambda_{2}} \log{(\lambda_{2}\nu)}
\quad , \quad 
f=c_{1}+c_{2}(\lambda_{2}\nu)^{-n}
\quad , \quad 
g=\frac{n}{(n-2)}\frac{1}{\lambda_{2}}[c_{1}+c_{2}(\lambda_{2}\nu)^{2-n}]
\,,
\end{equation}
and
\begin{equation}
\!\!\!\!\!
\begin{cases}
E=\sqrt{\frac{\nu}{\lambda_{2}}}
\qquad , \qquad f=c_{1}(\lambda_{2}\nu)^{\frac{1-n}{2}}H_{2F1}[\frac{1-n}{2},\frac{n}{2},\frac{3}{2},4\lambda_{2}\nu]
\,,
\\
g=\frac{1+4(n-1)\lambda_{2}\nu}{2(n-2)}Ef+c_{1}(\lambda_{2}\nu)^{\frac{4-n}{2}}\frac{2n(n-1)}{3\lambda_{2}(n-2)}(4\lambda_{2}\nu-1) H_{2F1}[\frac{3-n}{2},\frac{2+n}{2},\frac{5}{2},4\lambda_{2}\nu] \, .
\end{cases}
\end{equation}
Obviously, having fixed the interaction function to solve \eqref{system-first-form}, the above solutions will generically not satisfy SZ flows, but simply provide examples of conserved currents in AF single free boson theories defined by the given $E$.
\end{itemize}
The system \eqref{system-first-form} can be rewritten in a somewhat nicer form by introducing integrating factors. Respectively multiplying the equations by $a(\nu)$ and $b(\nu)$ of the form
\begin{equation}
a(\nu) :=c_{1}\nu^{-1}E''[1-(E')^2]^{-1-\tfrac{n}{2}}\,,
\qquad \qquad
b(\nu) := c_{2}\nu^{n-1}E''[1-(E')^2]^{\tfrac{n}{2}-2} \, ,
\end{equation}
and defining  new functions
\begin{equation}\label{F,Gdefinition}
F:= \frac{c_{1}}{n-2} [1-(E')^2]^{-\tfrac{n}{2}}f \,,\qquad \qquad G:=-\frac{c_{2}}{n}\nu^{n-2}[1-(E')^2]^{-1+\tfrac{n}{2}}g \, ,
\end{equation}
with $c_{1},c_{2}$ dimensionless constants, one ends up with
\begin{equation}\label{system-second-form}
\begin{aligned}
F'&=-\frac{c_{1}}{c_{2}}n\nu^{1-n}E''[1-(E')^2]^{-n}G \, , 
\\
G'&=\frac{c_{2}}{c_{1}}(n-2)\nu^{n-1}E''[1-(E')^2]^{n-2}F \, .
\end{aligned}
\end{equation}
The latter system has several interesting features:
\begin{itemize}
\item It is manifestly invariant under $n\rightarrow2-n$ and the relabelling of $F\leftrightarrow G$ and $c_{1}\leftrightarrow c_{2}$, which also leaves invariant the definition \eqref{F,Gdefinition} of $F,G$ if one further relabels $f\leftrightarrow g$. This encodes an exchange in role of the currents \eqref{free-bosono-higher-spin-currents-guess}, which should be expected already at the level of \eqref{system-original-form-appendix}, but which seems to be less evident.
\item The right hand side of both equations contains a total derivative due to the relation
\begin{equation}\label{relation1}
E''[1-(E')^2]^\alpha = \frac{\mathrm{d}}{\mathrm{d}\nu}\Bigl[ E'H_{2F1}[\tfrac{1}{2},-\alpha,\tfrac{3}{2},(E')^2]\Bigr] \quad \quad \forall \alpha \in \mathbb{R} \, .
\end{equation}
\item The equations in \eqref{system-second-form} can be combined in two nice ways by taking their product
\begin{equation}\label{product-of-derivatives}
\Bigl[\frac{\mathrm{d}}{\mathrm{d}\nu}\log{F}  \Bigr] \Bigl[\frac{\mathrm{d}}{\mathrm{d}\nu}\log{G}  \Bigr] 
= -n (n-2)\Bigl[\frac{\mathrm{d}}{\mathrm{d}\nu}\Bigl(\text{arctanh}(E') \Bigr) \Bigr]^2 \, ,
\end{equation}
and their ratio
\begin{equation}\label{ratios-of-derivatives}
\Bigl[\frac{\mathrm{d}}{\mathrm{d}\nu}F^2 \Bigr]=-\frac{n}{n-2}\Bigl[\frac{\mathrm{d}}{\mathrm{d}\nu}G^2 \Bigr] \Bigl(\nu[1-(E')^2]\Bigr)^{2(1-n)} \, .
\end{equation} 
\end{itemize}

\subsection*{Decoupled 2nd order ODEs and Volterra integral equation}
We conclude by observing that the coupled 1st order system \eqref{system-second-form} can be rewritten as two decoupled 2nd order ODEs for $F$ and $G$\footnote{The same can of course be done for the original system \eqref{system-first-form}, but the result is much more involved and devoid of any obvious symmetry.}:
\begin{equation}\label{2nd-order-ODEs-FG}
\begin{aligned}
F''(\nu)+F'(\nu) p(E',E'',n)+F(\nu)q(E',E'',n)&=0 \, ,
\\
G''(\nu)+G'(\nu) p(E',E'',2-n)+G(\nu)q(E',E'',2-n)&=0 \, ,
\end{aligned}
\end{equation}
where we defined
\begin{equation}\label{2nd-order-ODE-coefficients}
\begin{aligned}
p(E',E'',n)&:=\frac{\mathrm{d}}{\mathrm{d}\nu}\Bigl[ \log{\Bigl( \frac{\nu^{n-1}[1-(E')^2]^n}{E''}\Bigr) } \Bigr] \neq p(E',E'',2-n) \, , 
\\
q(E',E'',n)&:=\Bigl[\frac{\mathrm{d}}{\mathrm{d}\nu}\Bigl(\sqrt{n(n-2)}\,\text{arctanh} \,E' \Bigr) \Bigr]^2 = q(E',E'',2-n) \, .
\end{aligned}
\end{equation}
From this rewriting it is clear that $F$ and $G$ must satisfy exactly the same 2nd order ODE, the only difference being the replacement $n\rightarrow2-n$ in the coefficient functions (one of which is actually left invariant). This is the somewhat improved version of the symmetry noted in the first bullet point below equation \eqref{system-second-form} and explicitly confirms that the currents \eqref{free-bosono-higher-spin-currents-guess} only differ by the power of auxiliary fields contained in the trace. 
\begin{itemize}
\item The 2nd order ODEs seem to allow for an easier extraction of the currents in the presence of polynomial interaction functions.
For example, considering 
\begin{equation}\label{polynomial-order2-for-E}
E=\lambda_{0}+\lambda_{1}\nu + \lambda_{2}\nu^2 \, ,
\end{equation}
one finds
\begin{align}\label{solution-2nd-order-ODE}
&F= 2^{n-1}\Bigl(\frac{-1+\lambda_{1}+2\lambda_{2}\nu}{\lambda_{2}}\Bigr)^{-n/2}\Bigl(\frac{1+\lambda_{1}+2\lambda_{2}\nu}{\lambda_{2}}\Bigr)^{1-n/2}
\\
& \Bigl( c_{4}\text{HeunG}\Bigl[ \frac{\lambda_{1}-1}{\lambda_{1}+1},\frac{-(n-1)-(n-1)^2\lambda_{1}}{\lambda_{1}+1} ,1-n,-1+2n,-1+n,2,-\frac{2\lambda_{2}\nu}{\lambda_{1}+1}\Bigr]+
\notag \\
& + c_{5}\nu^{2-n}\text{HeunG}\Bigl[ \frac{\lambda_{1}-1}{\lambda_{1}+1},\frac{n-3-(n^2-3)\lambda_{1}}{\lambda_{1}+1} ,3-2n,1+n,3-n,2,-\frac{2\lambda_{2}\nu}{\lambda_{1}+1}\Bigr]\Bigr) \, ,
\notag
\end{align}
and $G$ is the same with $n\!\rightarrow\! 2\!-\!n$. From these it is straightforward to extract $f,g$ using the relations \eqref{F,Gdefinition}. 
Notice that from \eqref{dimension-summary1} and \eqref{F,Gdefinition} one finds
\begin{equation}
[F]=0
\quad, \quad
[G]=2(n-1)
\quad , \quad
[F']=-2
\quad , \quad
[G']=2(n-2) \, ,
\end{equation}
which is consistent with the equations \eqref{system-second-form} and the solution \eqref{solution-2nd-order-ODE} provided that $[c_{4}]=2(1-n)$ and $[c_{5}]=2(3-2n)$, or equivalently $c_{4}\propto\lambda_{2}^{n-1}$ and $c_{5}\propto\lambda_{2}^{2n-3}$.
Notice also that the choice of interaction function \eqref{polynomial-order2-for-E} simply represents a truncation of some analytic function $E(\nu)=\sum_{k=0}^{\infty}\epsilon_{k}\nu^{k}$ considered in section \ref{subsec:perturbative-analysis-free-boson}. 
\item The special form of the coefficient functions \eqref{2nd-order-ODE-coefficients} in \eqref{2nd-order-ODEs-FG} can be exploited to obtain an implicit integral characterisation of the system. Writing the $F$-equation as
\begin{equation}\label{intermediate-Volterra}
F''(\nu)+F'(\nu)P'(\nu)+F(\nu)Q'(\nu)^2 =0 \, ,
\end{equation}
with
\begin{equation}
P(\nu):=\log{\Bigl( \frac{\nu^{n-1}[1-(E')^2]^n}{E''}\Bigr) }\,,
\qquad \qquad
Q(\nu):= \sqrt{n(n-2)}\,\text{arctanh} \,E' \, ,
\end{equation}
and multiplying by the integrating factor
\begin{equation}
d(\nu):=e^{\frac{1}{2}P(\nu)} \, ,
\end{equation}
one can write \eqref{intermediate-Volterra} as a Schroedinger equation
\begin{equation}\label{schroedinger-eq}
\psi''(\nu)+V(\nu)\psi(\nu)=0 \, ,
\end{equation}
with
\begin{equation}
\psi(\nu):=F(\nu)e^{\frac{1}{2}P(\nu)}
\qquad \text{and} \qquad
V(\nu):=Q'(\nu)^2-\frac{1}{2}P''(\nu)-\frac{1}{4}P'(\nu)^2 \, .
\end{equation}
The explicit potential takes the following complicated form:
\begin{equation}
V(\nu)\!:=\!\tfrac{[(E')^2-1](E'')^2-8\nu E'(E'')^3-8\nu^2(E'')^4-3\nu^2[(E')^2-1](E''')^2+2\nu[(E')^2-1]E''(E'''+\nu E'''')}{4\nu^2[(E')^2-1](E'')^2} \, ,
\end{equation}
and given the initial conditions
\begin{align}
    \psi ( 0 ) = a \, , 
    \qquad \qquad 
    \psi' ( 0 ) = b \, ,
\end{align}
the general solution to \eqref{schroedinger-eq} can be expressed as a Volterra integral equation,
\begin{align}
    \psi ( \nu ) = a + b \nu - \int_0^\nu \mathrm{d}s \, \left( \nu - s \right) V ( s ) \psi ( s ) \, .
\end{align}
This is hard to solve because $\psi$ appears both on the left and under the integral on the right, but can be solved perturbatively for small $V$ by expanding around the solution
\begin{align}
    \psi_0 = a + b \nu \, ,
\end{align}
as a WKB-type expansion of the form
\begin{align}
    \psi ( \nu ) \!=\! \psi_0 ( \nu ) 
    \!-\!\!\! \int_0^\nu \!\!\!\!\mathrm{d}s  ( \nu \!-\! s ) V ( s ) \psi_0 ( s ) \!+\! \int_0^\nu \!\!\!\! \mathrm{d}s  ( \nu \!-\! s ) V ( s ) \!\! \int_0^s \!\!\!\!\mathrm{d}t  ( s \!-\! t ) V ( t ) \psi_0 ( t ) \!-\! ... \, .
\end{align}
This is further evidence, together with theorem \ref{nogo_thm}, that despite its seemingly simple structure the system of 1st order ODEs \eqref{system-original-form-appendix} does not allow for a nice closed-form solution for generic $n>2$. While various interesting transformations can be applied to the system, its complexity remains unchanged and at the end of the day the best one can hope for is obtaining perturbative solutions in the spirit of section \ref{subsec:perturbative-analysis-free-boson}.
\end{itemize}

\newpage

\section{Vanishing of the commutator terms}\label{appendix:identity}
In this appendix we show that given the conditions \eqref{AF-general-EOM-for-identity}, here reported for simplicity,
\begin{equation}\label{assumptions-appendix}
\begin{aligned}
D_{+}\mathcal{B}_{-}\!+\!D_{-}\mathcal{B}_{+} \deq & \, 0 
\quad\,\,\,\,\, \text{with} \quad\,\,\,\,\,
\mathcal{B}_{\pm} := -(\mathcal{A}_{\pm}+2v_{\pm}) \, ,
\\
\mathcal{A}_{\pm}+v_{\pm}+\Delta_{\pm} \deq & \, 0 
\quad\,\,\,\,\, \text{with} \quad\,\,\,\,\, 
\Delta_{\pm}:= \delta_{v_{\pm}}E(v) 
\quad \text{and} \quad [\Delta_{\pm},v_{\mp}]\deq \, 0 \, ,
\\
D_{+}\mathcal{A}_{-} \!-\!D_{-}\mathcal{A}_{+} \!+\! a \,[\mathcal{A}_{+},\mathcal{A}_{-}] \deq & \, 0 
\quad\,\,\,\,\, \text{with} \quad\,\,\,\,\, 
 D_{\pm}:=\partial_{\pm}\!+\![\mathcal{C}_{\pm},-]
\quad \text{and} \quad a \! \in \! \mathbb{R} \, \text{const.} 
\end{aligned}
\end{equation}
the commutator terms in the last step of \eqref{identity-steps} vanish, namely
\begin{equation}\label{desired-appendix}
\mathrm{tr}\Bigl(v_{\pm}^{n-1}[\mathcal{A}_{+},\mathcal{A}_{-}] \Bigr) \deq \, 0  
\qquad \qquad \text{and} \qquad \qquad  
\mathrm{tr}\Bigl(v_{\pm}^{n-1}[\mathcal{C}_{\pm},\Delta_{\mp}]\Bigr) \deq 0 \, .
\end{equation}
To begin we recall an identity, holding for any Lie-algebra valued $A,B,C$
\begin{equation}\label{Lie-algebra-identity}
\mathrm{tr}\Bigl( C^{n-1}[A,B] \Bigr) = \sum_{k=0}^{n-2}   \mathrm{tr} \Bigl( AC^{k}[B,C]C^{n-2-k} \Bigr)  \, ,
\end{equation}
which can be explicitly derived starting from
\begin{equation}
\mathrm{tr}\Bigl( C^{n-1}[A,B] \Bigr) = \mathrm{tr}\Bigl( ABC^{n-1} \Bigr) - \mathrm{tr}\Bigl( AC^{n-1}B \Bigr)
\end{equation}
and repeatedly commuting $B$ through $C^{n-1}$ in both terms on the right hand side. The second relation in \eqref{desired-appendix} is then readily verified by using \eqref{Lie-algebra-identity} with $A\!\equiv\! \mathcal{C_{\pm}}$, $B\!\equiv\! \Delta_{\mp}$, $C\!\equiv\! v_{\pm}$ and exploiting the commutator $[\Delta_{\pm},v_{\mp}]\deq 0$ in \eqref{assumptions-appendix}. The latter assumption also implies 
\begin{equation}\label{commutator[A,v]}
[\mathcal{A}_{\pm},v_{\mp}] \deq -[v_{\pm},v_{\mp}] \, ,
\end{equation}
which combined with \eqref{Lie-algebra-identity} for $A\!\equiv\! \mathcal{A}_{+}$, $B\!\equiv\! \mathcal{A}_{-}$, $C\!\equiv\! v_{+}$ brings the first relation in \eqref{desired-appendix} to
\begin{equation}
\mathrm{tr}\Bigl(v_{+}^{n-1}[\mathcal{A}_{+},\mathcal{A}_{-}] \Bigr) \!=\! \sum_{k=0}^{n-2}  \mathrm{tr} \Bigl( \mathcal{A}_{+}v_{+}^{k}[\mathcal{A}_{-},v_{+}]v_{+}^{n-2-k} \Bigr) \!\deq\!  - \! \sum_{k=0}^{n-2}   \mathrm{tr} \Bigl( \mathcal{A}_{+}v_{+}^{k}[v_{-},v_{+}]v_{+}^{n-2-k} \Bigr) \, .
\end{equation}
Using now \eqref{Lie-algebra-identity}, this time in the reversed direction with $A\!\equiv\! \mathcal{A}_{+}$, $B\!\equiv\! v_{-}$, $C\!\equiv\! v_{+}$, and exploiting once again \eqref{commutator[A,v]},  one finally arrives at
\begin{equation}
\begin{aligned}
\mathrm{tr}\Bigl(v_{+}^{n-1}[\mathcal{A}_{+},\mathcal{A}_{-}] \Bigr) \deq  - \mathrm{tr}\Bigl(v_{+}^{n-1}[\mathcal{A}_{+},v_{-}] \Bigr) \deq \mathrm{tr}\Bigl(v_{+}^{n-1}[v_{+},v_{-}] \Bigr) = 0 \, ,
\end{aligned}
\end{equation}
which is the desired first relation in \eqref{desired-appendix} for the choice of upper sign in $v_{\pm}$. The argument remains unchanged, up to an overall sign, for the choice of lower sign in $v_{\pm}$.

\newpage

\section{Derivative identities for $\mathfrak{su}(3)$}\label{app:su3der}
Here we collect the identities involving $\partial_{\pm}\mathrm{tr}(v_{\mp}^3)$ and $\partial_{\pm}\mathrm{tr}(v_{\mp}^2)$ derived using the relation \eqref{important-identity} and used in section \ref{sec:solution-spin-3-flow} to solve the spin-$3$ flow equation.
To begin, note that $\Delta_{\pm}:=\delta_{v_{\mp}}E(v) $ is linear in derivatives of $E$ along the Lorentz invariant variables $\nu _{2}, \, \nu_{3}, \, \omega,\, \widetilde \omega $ and hence we can compute the derivative identities using \eqref{important-identity} term by term, adding up all the contributions -- proportional to $\del _{\nu _{2}}E$, $\del _{\nu _{3}}E$, $\del _{\omega}E$ and $\del _{\widetilde \omega}E$ -- at the end. 

\subsection*{$\del _{\nu _3}E$ contributions}
We only present a detailed derivation of the $E_{3}:=\del _{\nu _3}E$ terms, since these rely on some specific properties of $\mathfrak {su }(3)$. The $E_{\nu _2}$, $E_{\omega} $ and $E_{\widetilde \omega }$ terms can be derived in a similar way. The equations of motion imply 
 \begin{equation}\label{rel-appendix-c}
    \begin{split}
        \frac 13 \del _- \, \ttr {v_+^3}&=  \del _+(3\, E_3\, \ttr{v_-^3}\, v_+{}^A\, v_+{}^B\,)v_+{}^C\, v_+{}^D\, d_{ABE}\, \ttr {T^E\, T_C\,T_D}+ \ldots
        \,,
    \end{split}
 \end{equation}
where the dots stand for omitted terms and we specialised the definitions in \eqref{d-symmetric-tensor} to
\begin{equation}
d_{ABC}:=\mathrm{tr}(T_{(A}T_{B}T_{C)}) \,.
\end{equation}
The indices $CD$ in \eqref{rel-appendix-c} are symmetrised, so we can in fact write 
\begin{equation}
     d_{ABE}\, \ttr{T^E\, T_C\,T_D} 
     \qquad 
     \rightarrow 
     \qquad 
     d_{ABE}\, d^E{}_{CD}
     \,,
\end{equation}
having used the Cartan-Killing form $\gamma_{AB}:=\mathrm{tr}(T_{A}T_{B})$ and its inverse $\gamma^{AB}$ to lower/raise indices as $d^{E}{}_{CD}:=\gamma^{EF}d_{FCD}$.
In addition, the derivative can only hit at most one $v_+$, so 3 out of the four indices will also be symmetrised, hence we really get 
\begin{equation}
    d_{ABE}\, d^E{}_{CD}
    \qquad \rightarrow \qquad 
    d_{E(AB}\, d^E{}_{CD)}= \frac 16 \gamma _{(AB}\gamma _{CD)}\,.
\end{equation}
The full expansion becomes
\begin{equation}
    \begin{split}
        \frac 13 \del _- \ttr {v_+^3}&= \frac12 \del _+ (E_3\, \tr(v_-^3))\ttr{v_+^2 }^2 +  E_3 \, \ttr{v_-^3}\,\del_+ v_+{}^A\, v_+{}^B\, v_+ {}^C\, v_+{}^D\gamma_{(AB}\,\gamma _{CD)}\\[0.6em]
        &= \frac 12 \del _+ (E_3\, \tr(v_-^3))\ttr{v_+^2 }^2 +\frac 12  E_3 \, \ttr{v_-^3}\,\del_+\ttr{v_+^2}\, \ttr{v_+^2}
        \,,
    \end{split}
\end{equation}
and flipping signs gives us 
\begin{equation}\label{eq:E3terms}
\frac 13 \del _{\mp} \, \ttr{v_{\pm}^3} =\frac 12 \del _{\pm} (E_3\, \tr(v_{\mp}^3))\ttr{v_{\pm}^2 }^2 + \frac12 E_3 \, \ttr{v_{\mp}^3}\,\del_{\pm}\ttr{v_{\pm}^2}\, \ttr{v_{\pm}^2}\,.
\end{equation}
The analogous expressions for $\ttr {v_{\pm}^2}$ are simpler 
\begin{equation}
    \begin{split}
        \frac 12 \del _- \, \ttr{v_+^2} &= \tr \left (\del _+(3 E_3\ttr{v_-^3}\,v_+{}^A\, v_+{}^B)T^C \, v_+\right)\, d_{ABC}\\
        &= 3 \del _+\left(E_3\tr (v_-^3)\right)\tr (v_+^3)+ 2E_3\tr (v_-^3)\del _+\tr (v_+^3)\,,
    \end{split}
\end{equation}
and finally 
\begin{equation}
\frac 12 \del _{\mp} \, \tr (v_{\pm}^2) =3 \del _{\pm}\left(E_3\tr (v_{\mp}^3)\right)\tr (v_{\pm}^3)+ 2E_3\tr (v_{\mp}^3)\del _{\pm}\tr (v_{\pm}^3) \, .
\end{equation}

\subsection*{Full derivative identities}
Upon including $\omega $ and $\widetilde \omega $ in the deformation function, and after adding the $E_{2}$ and $E_{3}$ contributions, we obtain the following relations
\allowdisplaybreaks
\begin{align}\label{eq:su3derid}
        \frac 13 \del _-\ttr {v_+^3}&= 2 \del_+\rb {E_2\ttr{v_-^2}}\ttr {v_+^3}+ \frac 23 E_2\ttr {v_-^2}\del _+ \ttr {v_+^3}+\frac12 \del _+ (E_3\, \tr(v_-^3))\tr (v_+^2 )^2 
        \nonumber \\
        &+ \frac12 E_3 \, \tr (v_-^3)\,\del_+\tr (v_+^2)\, \tr (v_+^2)+ \del _+\left (E _{ \omega}\,\ttr {v_-^2}^3\, \ttr {v_+^3} \right )\ttr{v_+^2}^2 
        \nonumber \\
        &+ E_\omega \,\ttr {v_-^2}^3\, \ttr {v_+^3}\del _+\ttr{v_+^2}\ttr{v_+^2}+6\del_+\left (E_{\widetilde \omega } \ttr{v_-^3}^2\ttr{v_+^2}^2\right )\ttr{v_+^3} 
        \nonumber \\
        &+2E_{\widetilde \omega } \ttr{v_-^3}^2\ttr{v_+^2}^2\del _+\ttr {v_+^3}\,,
        \\[1em]
        \frac 13 \del _+\ttr {v_-^3}&= 2 \del_-\rb {E_2\ttr{v_+^2}}\ttr {v_-^3}+ \frac 23 E_2\ttr {v_+^2}\del _- \ttr {v_-^3}+\frac12\del _- (E_3\, \tr(v_+^3))\tr (v_-^2 )^2 
        \nonumber  \\
        &+ \frac12E_3 \, \tr (v_+^3)\,\del_-\tr (v_-^2)\, \tr (v_-^2)+ \del _-\left (E _{ \widetilde\omega}\,\ttr {v_+^2}^3\, \ttr {v_-^3} \right )\ttr{v_-^2}^2 
        \nonumber \\
        &+ E_{\widetilde \omega} \,\ttr {v_+^2}^3\, \ttr {v_-^3}\del _-\ttr{v_-^2}\ttr{v_-^2}+6\del_-\left (E_{ \omega } \ttr{v_+^3}^2\ttr{v_-^2}^2\right )\ttr{v_-^3} 
        \nonumber \\
        &+2E_{ \omega } \ttr{v_+^3}^2\ttr{v_-^2}^2\del _-\ttr {v_-^3},
        \\[1em]
        \frac 12 \del_- \ttr {v_+^2}&= 6 \del _+ \rb {E_\omega \ttr {v_-^2}^3\ttr {v_+^3}}\ttr {v_+^3}+ 4 E_{\omega}\ttr {v_-^2}^3\ttr {v_+^3}\del _+ \ttr {v_+^3} 
        \nonumber \\
        &+6\del _+ \rb {E_{\widetilde \omega}\ttr{v_-^3}^2\ttr {v_+^2}^2}\ttr {v_+^2}+3E_{\widetilde \omega}\ttr {v_-^3}^2\ttr {v_+^2}^2\del _+ \ttr {v_+^2} 
        \nonumber \\
        &+3 \del _+\left(E_3\tr (v_-^3)\right)\tr (v_+^3)+ 2E_3\tr (v_-^3)\del _+\tr (v_+^3) 
        \nonumber \\
        &+2\del _+ \rb{E_2\ttr{v_-^2}}\ttr{v_+^2} + E_2\ttr {v_-^2}\del _+\ttr {v_+^2}\,, 
        \\[1em]
        \frac 12 \del_+ \ttr {v_-^2}
        &= 6 \del _- \rb {E_{\widetilde \omega} \ttr {v_+^2}^3\ttr {v_-^3}}\ttr {v_-^3}+ 4 E_{\widetilde \omega}\ttr {v_+^2}^3\ttr {v_-^3}\del _- \ttr {v_-^3} 
        \nonumber \\
        &+6\del _- \rb {E_{ \omega}\ttr{v_+^3}^2\ttr {v_-^2}^2}\ttr {v_-^2}+3E_{ \omega}\ttr {v_+^3}^2\ttr {v_-^2}^2\del _- \ttr {v_-^2} 
        \nonumber \\
        &+3 \del _-\left(E_3\tr (v_+^3)\right)\tr (v_-^3)+ 2E_3\tr (v_+^3)\del _-\tr (v_-^3) 
        \nonumber \\
        &+2\del _- \rb{E_2\ttr{v_+^2}}\ttr{v_-^2} + E_2\ttr {v_+^2}\del _-\ttr {v_-^2}\,.
\end{align}

\bibliographystyle{utphys}
\bibliography{master}
\end{document}